%% file: bare_jrnl.tex
\documentclass[journal,12pt,onecolumn,draftclsnofoot]{IEEEtran}
\usepackage{amsfonts}
\usepackage{graphicx}
\usepackage{booktabs}
\usepackage{algorithm}
\usepackage{algorithmic}
\usepackage{graphics}
\usepackage{subfigure}
\usepackage{multirow}
\usepackage{cite}
\usepackage{bigstrut,multirow,rotating}
\usepackage{amsmath,amsthm,amssymb}
\usepackage{array,diagbox,siunitx}
\usepackage{mathrsfs}
\usepackage{bm}
\usepackage{color}
\usepackage{setspace}

\usepackage{xcolor}
\usepackage{xpatch}

\ifCLASSINFOpdf
\else
\fi
\begin{document}
\title{Deep CSI Compression for Massive MIMO: A Self-information Model-driven Neural Network}

\author{Ziqing Yin,
        Wei Xu,~\IEEEmembership{Senior Member,~IEEE},
        Renjie Xie,
        Shaoqing Zhang,\\
        Derrick Wing Kwan Ng,~\IEEEmembership{Fellow,~IEEE},
        and~Xiaohu You,~\IEEEmembership{Fellow,~IEEE}% <-this % stops a space
\thanks{Z. Yin, W. Xu, R. Xie, S. Zhang, and X. You are with the National Mobile Communications Research Laboratory, Southeast University, Nanjing 210096, China, and also with the Purple Mountain Laboratories, Southeast University, Nanjing 210096, China (e-mail: zqyin@seu.edu.cn, wxu@seu.edu.cn, renjie\_xie@seu.edu.cn, sq\_zhang@seu.edu.cn, xhyou@seu.edu.cn).}% <-this % stops a space
\thanks{D. W. K. Ng is with the School of Electrical Engineering and Telecommunications, University of New South Wales, Sydney, NSW 2052, Australia (e-mail: w.k.ng@unsw.edu.au).}}%

\maketitle

\begin{abstract}
In order to fully exploit the advantages of massive multiple-input multiple-output (mMIMO), it is critical for the transmitter to accurately acquire the channel state information (CSI). Deep learning (DL)-based methods have been proposed for CSI compression and feedback to the transmitter. Although most existing DL-based methods consider the CSI matrix as an image, structural features of the CSI image are rarely exploited in neural network design. As such, we propose a model of self-information that dynamically measures the amount of information contained in each patch of a CSI image from the perspective of structural features. Then, by applying the self-information model, we propose a model-and-data-driven network for CSI compression and feedback, namely IdasNet. The IdasNet includes the design of a module of self-information deletion and selection (IDAS), an encoder of informative feature compression (IFC), and a decoder of informative feature recovery (IFR). In particular, the model-driven module of IDAS pre-compresses the CSI image by removing informative redundancy in terms of the self-information. The encoder of IFC then conducts feature compression to the pre-compressed CSI image and generates a feature codeword which contains two components, i.e., codeword values and position indices of the codeword values. Subsequently, the IFR decoder decouples the codeword values as well as position indices to recover the CSI image. Experimental results verify that the proposed IdasNet noticeably outperforms existing DL-based networks under various compression ratios while it has the number of network parameters reduced by orders-of-magnitude compared with various existing methods.

\newpage

\end{abstract}


% \fi
\IEEEpeerreviewmaketitle

\input{sections/01_introduction}
\input{sections/02_system}
\input{sections/03_model}

\input{sections/04_experiments}

\input{sections/05_conclusion}
\ifCLASSOPTIONcaptionsoff
  \newpage
\fi



\bibliographystyle{IEEEtran}

\bibliography{IEEEabrv.bib}

% IEEEabrv.bib
% \end{thebibliography}


% \end{IEEEbiography}


% % insert where needed to balance the two columns on the last page with
% %\newpage



%\vfill


\end{document}

%% file: sections/01_introduction.tex
\section{Introduction}
\IEEEPARstart{W}{ith} the development of the fifth-generation (5G) wireless communication networks, massive multiple-input multiple-output (mMIMO) has become a key technology \cite{9205230}. By deploying a large number of antennas, mMIMO not only improves the channel capacity greatly with limited spectral resources, but also has a strong ability of multiuser interference suppression \cite{6994239}. In order to reap the advantages of mMIMO, the transmitter needs to obtain accurate channel state information (CSI) of the channel. Recently, intensive researches on mMIMO have been conducted in the fields of channel estimation \cite{4357450} and channel compression feedback \cite{8972904,9445070}. In particular, in frequency division duplexed (FDD) systems, user equipments (UEs) estimate the CSI of downlink channels and then feed the CSI back to the base station (BS) through a dedicated feedback link with limited bandwidth. However, the overhead of CSI feedback becomes enormous due to the increasing number of antennas of mMIMO, which makes the design of efficient CSI feedback challenging. 

To reduce the feedback signaling overhead of CSI feedback, researchers have devised numerous algorithms by using various estimation and compression theories. In particular, most studies aimed to reduce the signaling overhead by exploiting spatial and temporal correlations of  mMIMO channels. For instance, an effective CSI feedback scheme was proposed in \cite{6214417} by applying compressed sensing (CS). The channel vector was compressed into a codeword  with reduced dimension by projecting it onto a sparse-basis at the UE. To facilitate practical implementation, this sparse-basis was chosen as popular orthogonal matrices, e.g., two-dimensional discrete Fourier transform (2D-DFT) matrix and two-dimensional discrete cosine transform (2D-DCT) matrix. Given that the basis was deterministic and known to all nodes, the CSI could be reconstructed from the codeword at the BS. Besides, an improved method was then proposed in \cite{6816089} by using a distributed compressive CSI estimation and the BS recovered the CSI matrix by exploiting a joint orthogonal matching pursuit 
recovery algorithm. Also in \cite{7166317}, an antenna  grouping-based method was proposed for further reducing the burden of CSI feedback in an FDD-based mMIMO system. Specifically, the proposed method in \cite{7166317}, namely antenna group beamforming (AGB), mapped multiple correlated antenna elements to a single representative value using predesigned patterns. However, the required feedback signaling overhead of these existing approaches is still exceedingly large  with the growing number of BS antennas in mMIMO systems, because the overhead scales linearly with the number of antennas, which limits their practicality.

Another line of works focused on designing vector and matrix codebooks for CSI compression. In fact, some pre-defined codebooks have been widely used for the CSI feedback in various commercial systems, e.g., LTE/LTE-A, IEEE 802.11n/ac, and WiMAX \cite{4641946}. In \cite{7434506}, a codebook was developed based on the theory of CS, which quantized low-dimensional channel measurements assuming no inter-cell interference. In \cite{4411709}, a codebook was designed with a reduced size by considering dominant line-of-sight (LoS) components between the UEs and the BS. As for handling non-LoS components, a rotated codebook based on channel statistics was proposed in \cite{1608648} for spatially correlated channels. Also, in \cite{8392731}, a channel subspace codebook was designed for CSI feedback by exploiting the knowledge of the angle-of-departure (AoD) of channels. Within a coherence time of the angular CSI, i.e., the AoD of mMIMO channels, the subspace codebook was able to quantize the channel vector accurately. In general, the design of codebooks is a sophisticated nonlinear procedure and specific codebooks need to be tailored for different types of channel distribution under various compression ratios.

\subsection{Related Work}
To overcome the challenges of conventional CSI compression methods and codebook designs, deep learning (DL)-based \cite{7298594} methods are getting attractive and appealing as promising alternatives. Due to the strong abilities of parallel calculation, adaptive learning, and cross-domain knowledge sharing, DL has been widely and successfully applied in areas of computer 
vision \cite{7780459}, speech recognition \cite{6296526}, and natural language process \cite{6737243}. Recently, it has started to draw increasing attention in the field of wireless communication. For instance, in \cite{9024047}, a data-driven DL network and a model-driven DL network were, respectively, proposed for channel estimation and signal 
detection for an uplink multiuser MIMO system. In \cite{8618345}, a DL-based framework was proposed for hybrid precoding design, where the deep neural network (DNN) was trained as a mapping function from CSI input to hybrid precoders.

To further unlock the potential of DL, it has also been introduced in the design of CSI compression and feedback \cite{8322184,9090892,8543184,9296555,9171358,9279228,9347820,9149229,9439959,9497358}. In \cite{8322184}, a DL network, named CsiNet, was proposed for CSI compression and feedback. Specifically by transforming the CSI matrix to an image representation, the UE adopted an encoder network to compress the CSI into a specific codeword for effective feedback. Then, the BS exploited a decoder network to recover the CSI image from the received codeword. In \cite{9090892}, a neural network named CQNet was proposed to jointly tackle CSI compression, codeword quantization, and recovery under a bandwidth constraint. By further considering the temporal correlation of wireless channels, an improved neural network was proposed in \cite{8543184}, which invoked a module, known as long short-term memory (LSTM) in both the encoder and the decoder networks. In particular, the module of LSTM helped  catch both temporal and frequency correlations of wireless channels. In \cite{9296555}, a DL-based CSI compression scheme called DeepCMC was proposed to improve the feedback performance by incorporating quantization
and entropy coding blocks.

For practical applications, considering that the obtained CSI is always noisy even at receiver, an anti-noise CSI compression network was proposed in \cite{9171358} by taking noisy CSI into consideration before CSI compression. By considering the following beamforming performance, a DL-based CSI feedback framework  was proposed in \cite{9279228} to maximize the ultimate goal of beamforming performance gain rather than the feedback accuracy. Also in \cite{9347820}, a joint neural network design of pilots and CSI estimate was proposed to improve the system performance. In particular, the neural network directly mapped the received pilots into a sequence of feedback bits at UEs and then the feedback bits from all the UEs were mapped directly into a precoding matrix at the BS by a neural network. To improve the accuracy of CSI feedback, a DL-based network named CRNet \cite{9149229} was proposed to achieve better performance via extracting CSI features on multiple resolutions. Besides, in \cite{9439959}, a neural network named ENet was trained for only the real part of CSI by exploiting the inherent correlation characteristics between the real and imaginary parts of complex-valued channel responses. Also in \cite{9497358}, a neural network named CLNet was proposed to utilize a forged complex-valued input layer to process signals and the spatial-attention to enhance the performance. These above DL-based approaches considered the CSI matrix as an image and optionally exploited the temporal and spatial channel correlations. However, they rarely consider the structural features, i.e., the shape and texture properties, of an image about the CSI. Moreover, these methods are mostly data-driven design of DL networks which requires a large number of data samples for training a network with a huge number of parameters, especially for applications in mMIMO. As a result, there is a need to design a lightweighted neural network for more effective CSI compression in FDD mMIMO systems.

Different from these existing methods using DL, this paper aims to consider the structural  features of CSI images from the perspective of information theory \cite{6773024} as well as image processing \cite{Carlucci_2019_CVPR}. In particular, by observing the image structural features, the CSI image is divided into shape patches and texture patches. In general, a shape patch is significantly different from its neighboring patches and contains essential information about the image. In contrast, a texture patch tends to repeat itself with slight and smooth changes in the neighboring region and it can be reconstructed easily from the neighboring patches in the process of CSI reconstruction. Thus, we conclude that the texture patch contains trivial information for the CSI reconstruction. We define this trivial information contained in the texture patch as the $\emph{informative redundancy}$. On the other hand, for CSI feedback with limited resources, it is natural to achieve accurate CSI reconstruction if more essential information, rather than informative redundancy, is contained in the feedback codeword. Therefore, removing informative redundancy in the CSI image can be beneficial for achieving better CSI compression and reducing the feedback signaling overhead.

% , we propose a definition of self-information, which is a target to measure the amount of information in the CSI image and reflects more directly the amount of information of each pixel in the CSI image. In this paper, we divide CSI matrix into several patches. If the patch contains more texture than shape, which means the patch is almost same with neighborhood patches, then the patch has a high likelihood and contains low self-information. On the contrary, if the patch contains more shape than texture, which means the patch is great different with neighborhood patches, then the patch has a low likelihood and contains high self-information. The patches with high self-information contain more "essential" information, which is more helpful for CSI reconstruction. The patches with low self-information contain more "insignificant" information, which is less helpful for CSI reconstruction. The insignificant information is also called redundancy, respectively. Therefore, removing redundancy effectively is necessary to achieve better performance of CSI reconstruction.

Considering the architecture of pure data-driven based DL networks, we motivate an integration of model-driven and data-driven designs of the network for efficient CSI compression with significantly reduced network complexity. Note that the model-driven component of the proposed network retains advantages of conventional  model-based iterative methods, such that it can exploit $a$ $priori$ information to enable a network with fewer trainable parameters and achieve convergence using a small set of training data samples. The CSI image pre-compressed by the model-driven network is used as the input of a subsequent data-driven neural network for further feature compression. This architecture of network integration reduces the required complexity of the data-driven network and accelerates the convergence of the entire network training.

% , which achieves more accurate compression to the CSI image compared with only data-driven networks \cite{2021Fully,8322184,8543184,9171358,9347820}. The model-driven network can remove redundancy efficiently by exploiting a prior knowledge based on the  structural feature of the CSI image. Then the data-driven network compresses the CSI image without redundancy to the codeword at UEs and recover the CSI image at the BS.

% In model-driven part, the network can remove redundancy effectively, which can also be regard as the pre-compression for CSI image. In data-driven part, the network stores the codewords and their corresponding position index simultaneously in the process of CSI compression based on CSI image without redundancy. Then the codewords and their corresponding position index are fed back to the decoder at the BS through a feedback link. The proposed network achieves better CSI reconstruction performance compared with only feed back codewords. In conclusion, in the model-driven network, the CSI image can effectively remove redundancy, which facilitates the subsequent compression storage. In data-driven network, The compression codewords and their corresponding position index acquired by encoder are fed back to the BS, and greatly improves the accuracy of CSI reconstruction.

\subsection{Contributions}
In this paper, we propose a DL-based CSI compression and feedback network, namely IdasNet, by exploiting the concept of image compression from the perspective of information theory. Different from directly compressing the CSI image via neural networks in existing methods, the IdasNet first pre-compresses the CSI image from the sense of self-information, i.e., removing the informative redundancy, then performs informative feature compression and informative feature decompression for the pre-compressed image. The main contributions of this paper are summarized as follows.

\begin{itemize}
\item By considering the structural features in terms of texture and shape of the CSI image, we propose a model of self-information, which measures the amount of information in a CSI image, and introduce a dynamic evaluation of the self-information in a patch-by-patch manner. Based on the evaluation of the self-information, we design a model-driven self-information deletion and selection (IDAS) module, which removes the informative redundancy of the original CSI image and outputs a series of selected image features for further compression. It is also verified experimentally that the feedback codeword acquired from the CSI image without informative redundancy contains more essential information than the codeword directly obtained from the original CSI image, which enhances the accuracy of CSI reconstruction.

% By applying the evaluation model of the self-information to the CSI image input, we design a new model-and-data-driven network, namely self-information deletion and selection network (IdasNet), for CSI compression and feedback. Different from directly compressing the CSI image via neural networks in existing methods, the IdasNet first pre-processes the CSI image from the sense of the self-information by a newly designed component of informative model-driven network. This informative model-driven network pre-compresses, i.e., removing the informative redundancy, the CSI image and outputs a series of selected image features for further compression. It is also verified experimentally that the feedback codeword, acquired from the CSI image without informative redundancy, contains more essential information than the codeword directly obtained from the original CSI image, which reduces the error of CSI reconstruction.

\item For the pre-compressed CSI image, pixels with larger self-information contain more channel information. In order for the codeword to carry more channel information, we design a data-driven compression module and a data-driven decompression module from the perspective of self-information. We design an Encoder network of informative feature compression (IFC) to generate the codeword which consists of codeword values and position indices of the codeword values. Correspondingly, a Decoder network of informative feature recovery (IFR) decouples the codeword values and position indices, which can dramatically improve the accuracy of CSI reconstruction at the BS.

% Given the informatively pre-compressed CSI features, we design an Encoder network of informative feature compression (IFC), rather than original CSI image compression, to generate a codeword of CSI features to feedback. Correspondingly, we also design a Decoder network of informative feature recovery (IFR) for recovering the CSI image from the codeword.

\item Experimental results verify that the proposed IdasNet outperforms existing DL-based networks for CSI feedback in terms of both recovery accuracy and network complexity. The proposed IdasNet achieves a performance gain of $3$ dB under different compression ratios in terms of normalized mean-squared error (NMSE) compared to existing methods. In addition, the IdasNet has a number of trainable parameters reduced by orders-of-magnitudes compared to the existing methods.
\end{itemize}

\subsection{Paper Organization and Notations}
The remainder of this paper is organized as follows. Section \uppercase\expandafter{\romannumeral2} introduces the system model. Section \uppercase\expandafter{\romannumeral3} 
proposes the definition and calculation of the self-information. Section \uppercase\expandafter{\romannumeral4} develops the DL-based network named IdasNet and elaborates the design details. Section \uppercase\expandafter{\romannumeral5} presents the simulation results. Conclusions are drawn in 
Section \uppercase\expandafter{\romannumeral6}. 

Throughout this paper, normal-face letters denote scalar variables, and boldface lower and uppercase symbols denote column vectors and matrices, respectively. The real part and the imaginary part of a complex matrix $\textbf{C}$ are denoted by $\mathcal{R}(\textbf{C})$ and $\mathcal{I}(\textbf{C})$, respectively. The superscript $(\cdot)^H$ denotes  Hermitian transpose of a matrix, respectively. Notation  $E\left \{ \cdot \right \}$ is the expectation operator, and $\mathbb{C}^{m \times n}$ represents the complex space of $m \times n$ dimensional matrices. Operator ${\parallel \cdot \parallel}_2$ returns the Euclidean norm.

%% file: sections/02_system.tex
\section{System Model}

\begin{figure*}[t]

\centering
\includegraphics[scale=0.37]{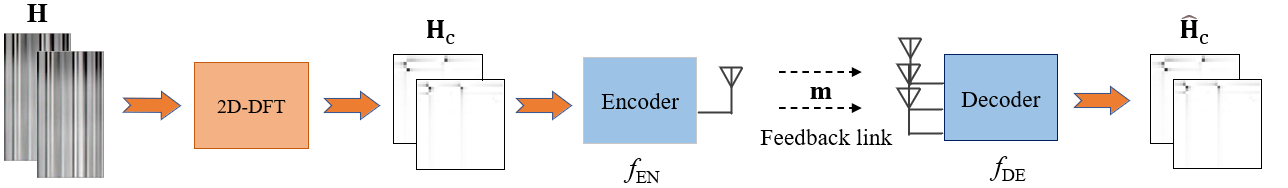}
\caption{A framework of typical DL network design for CSI compression and feedback.}
\label{fig:label1}
\end{figure*}
%%%%%%%%%经典深度学习网络结构图
We consider the downlink of an FDD mMIMO system, where $N_\text{r}$ antennas are deployed at the BS and a single-antenna is deployed at the UE. The system adopts
orthogonal frequency division multiplexing (OFDM) with $N_\text{s}$ subcarriers. The received signal at the $n$th subcarrier is presented as
\begin{equation}
    y_n = \textbf{h}_n^H \textbf{v}_n x_n + g_n\ \ n = 1,\ldots,N_\text{s},
\end{equation}
\noindent where $\textbf{h}_n \in \mathbb{C}^{N_\text{r} \times 1}$ denotes the channel vector at the $n$th subcarrier, $\textbf{v}_n \in \mathbb{C}^{N_\text{r} \times 1}$ is the corresponding precoding vector, $x_n \in \mathbb{C}$ is the transmitting signal, and $g_n \in \mathbb{C}$ is the
additive noise. The downlink channel matrix at all $N_\text{s}$ subcarriers is represented by $\textbf{H} = [\textbf{h}_1 \cdot \cdot \cdot \textbf{h}_{N_\text{s}}]^H$, whose size is $N_\text{s} \times N_\text{r}$. In FDD mMIMO systems, an estimate of $\textbf{h}_n$ is acquired at the UE and then this channel vector is quantized as a codeword by using a specific codebook. The obtained codeword is fed back to the BS through a limited feedback link. As such, the BS can reconstruct the CSI from the feedback codeword and then accordingly design the precoding vector $ \textbf{v}_n$ \cite{zsq}. 

In the spatial domain, the total number of feedback parameters, i.e., the size of $\textbf{H}$ as $N_\text{s} \times N_\text{r}$, is exceedingly large to be fed back with limited bandwidth. To facilitate the CSI compression, we transform the channel estimate at the UE from the spatial domain to the angular-delay domain \cite{1033686}. By applying the 2D DFT, the channel $\textbf{H}$ is transformed to the angular-delay domain as
\begin{equation}
    \textbf{H}_\text{a} = \textbf{F}_\text{c} \textbf{H} \textbf{F}_\text{d},
\end{equation}

\noindent where $\textbf{F}_\text{c} \in \mathbb{C}^{N_\text{s} \times N_\text{s}}$ and $\textbf{F}_\text{d} \in \mathbb{C}^{N_\text{r} \times N_\text{r}}$ denote the DFT matrices with corresponding sizes. Based on the fact that multipaths arrive at limited delay intervals \cite{8845636}, the channel $\textbf{H}_\text{a}$  contains nonzero values only in a small delay duration. Without loss of generality and following the same approach  in \cite{8322184}, we select the first $N_\text{c}$ rows of $\textbf{H}_\text{a}$, denoted by $\textbf{H}_\text{c}$, and the size of $\textbf{H}_\text{c}$ is $N_\text{c} \times N_\text{r}$. In this way, the number of parameters of feedback decreases from $N_\text{s} \times N_\text{r}$ to $N_\text{c} \times N_\text{r}$ in the angular-delay domain.  

In order to further reduce the feedback signaling overhead and acquire accurate CSI recovery at the BS, DL is applied for the CSI compression. The framework of a typical DL network for CSI feedback is shown in Fig. 1. An $Encoder$ network is deployed at the UE and it compresses the CSI image into a codeword of a specific dimension, which can be represented by
\begin{equation}
    \textbf{m} = f_{\rm{EN}}(\mathcal{R}(\textbf{H}_\text{c}), \mathcal{I}(\textbf{H}_\text{c})),
\end{equation}
\noindent where $f_{\rm{EN}}(\cdot)$ represents the compression function of the encoder. In particular, the encoder compresses the CSI image through $f_{\rm{EN}}$ to an $M$-dimensional vector $\textbf{m} \in \mathbb{R}^{M \times 1}$, where in general $M \ll N_\text{c} \times N_\text{r}$. Then the codeword, $\textbf{m}$, is fed back to the BS through a feedback link. In Fig. 1, the $Decoder$ network deployed at the BS recovers the CSI image from  $\textbf{m}$. The decoder network is represented by
\begin{equation}
    \widehat{\textbf{H}}_\text{c} = f_{\rm{DE}}(\textbf{m}),
\end{equation}
\noindent where $f_{\rm{DE}}(\cdot)$ represents the decompression function of the decoder, and $\widehat{\textbf{H}}_\text{c}$ is the recovered image of CSI. Note that the desired channel matrix $\widehat{\textbf{H}}$ in the spatial domain can be directly acquired by applying an inverse DFT to $\widehat{\textbf{H}}_\text{c}$. 

%% file: sections/03_model.tex
\section{Self-information of CSI Image}
In this section, by considering the structural features of the CSI image, we first introduce the definition of the self-information. The self-information is used to measure the amount of information contained in pixels of a CSI image. Inspired by the concept of informative dropout \cite{shi2020informative}, we propose a model of the self-information, which includes the estimate of the probability and the calculation of the self-information matrix.

\begin{figure*}[t]
\centering
\subfigure[Manhattan radius $R=1$.]{
\includegraphics[width = 3.6cm]{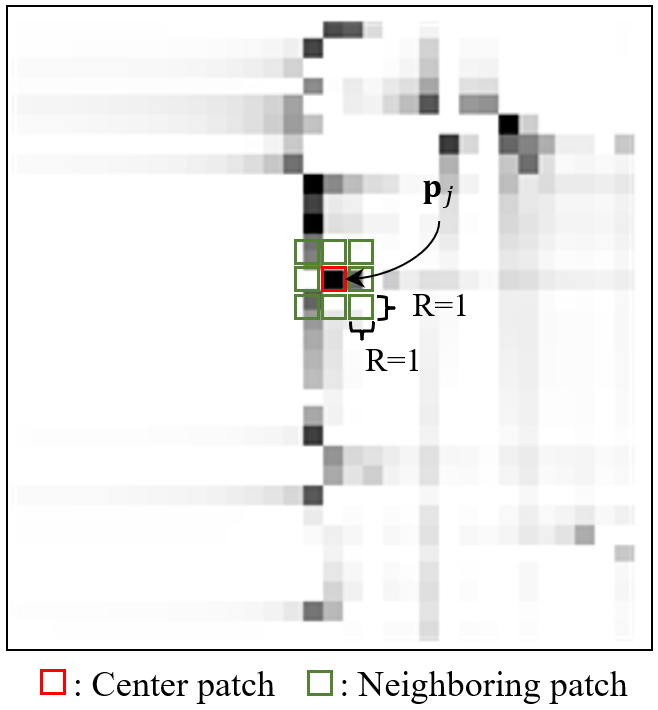}}
\subfigure[Manhattan radius $R=3$.]{
\includegraphics[width = 3.6cm]{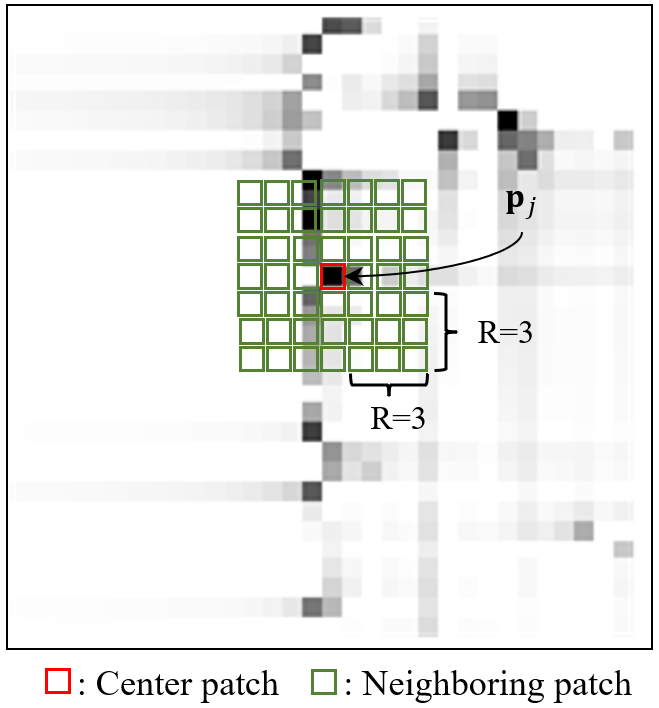}}
\subfigure[Original image.]{
\includegraphics[width = 3.6cm]{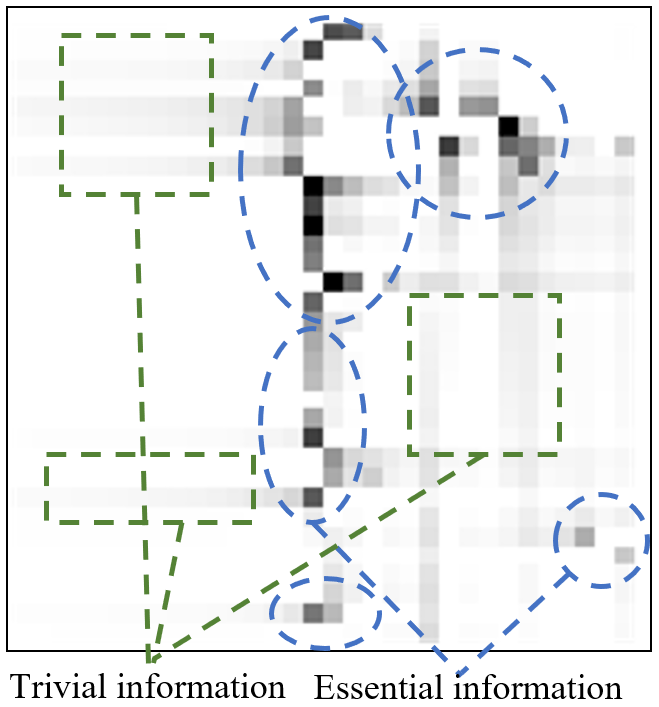}}
\subfigure[Self-information image.]{
\includegraphics[width = 3.6cm]{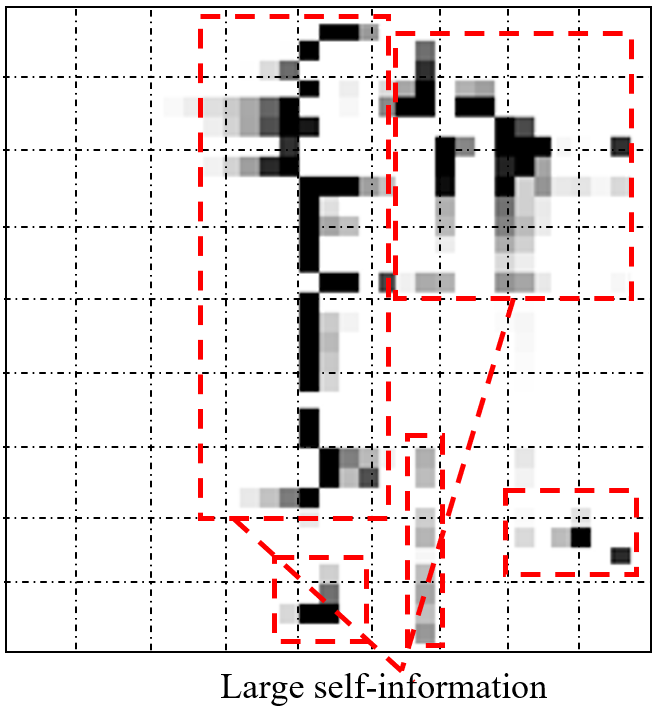}}
\caption{(a) Range of the neighboring patches set for $R=1$. (b) Range of the neighboring patches set for $R=3$. (c) The real part $\mathcal{R}(\textbf{H}_\text{c})$ of the original CSI image $\textbf{H}_\text{c}$. (d) The real part $\mathcal{R}(\textbf{H}_\text{c})$ in terms of self-information.}
\end{figure*}

\subsection{Definition of Self-information}
The definition of the self-information of the CSI image is inspired from the theory of image processing in computer vision. For an image, without loss of generality, we can divide it into some patches. Each patch is denoted by $\textbf{p}_j \in \mathbb{R}^{n \times n}$, where $n$ is the size of dividing grid and $j=\left \{1,2,\ldots, (N_\text{c} - n +1)(N_\text{r} - n +1) \right \}$. For an ordinary image, the $j$th patch, $\textbf{p}_j$, contains little information of this image if it contains mostly texture rather than shape, which means that $\textbf{p}_j$ looks almost the same as its neighboring patches. Otherwise, it is more informative if $\textbf{p}_j$ contains more shape than texture, which means that $\textbf{p}_j$ looks sharply different from its neighboring patches. To measure the amount of such information of $\textbf{p}_j$, we define the notion of self-information, denoted by $I_j$, by borrowing the concept of Shannon's work \cite{6773024}
\begin{equation}
    I_j = - \log_2 q_j, \ \ \forall j,
\end{equation}
\noindent where $\log{_2}$ is the base-$2$ logarithm and $q_j$ denotes the probability of $\textbf{p}_j$. Once the probability $q_j$ is obtained, we can calculate the corresponding self-information value $I_j$ to measure the amount of information contained in $\textbf{p}_j$. Note that $\textbf{p}_j$ with low probability contains a large amount of self-information and vice versa.

% \begin{figure}[t]
% \centering
% \subfigure[Original image.]{
% \includegraphics[width = 0.48\linewidth]{figures/f21.png}}
% \subfigure[Self-information image.]{
% \includegraphics[width = 0.48\linewidth]{figures/f22.png}}

% \caption{(a) The real part $\mathcal{R}(\textbf{H}_\text{c})$ of the original CSI image $\textbf{H}_\text{c}$. (b) The real part $\mathcal{R}(\textbf{H}_\text{c})$ in terms of self-information.}
% \label{fig:label2}
% \end{figure}

Specifically for a CSI image of interest, if $\textbf{p}_j$ has a large amount of self-information, equivalently low probability, then $\textbf{p}_j$ contains  essential information and is significantly helpful for CSI reconstruction. If $\textbf{p}_j$ has a tiny amount of self-information, equivalently large probability, then $\textbf{p}_j$ contains trivial information and contributes little for the CSI reconstruction. Intuitively, the trivial information corresponds to informative redundancy that can be removed with priority given a limited compression ratio. Removing the informative redundancy efficiently in CSI image ensures that the obtained codeword, $\textbf{m}$, can contain more essential information and thus increases the accuracy of CSI reconstruction.

\subsection{Calculation of Self-information}

In order to calculate the self-information of a CSI image, we first need to acquire the probability in terms of $q_j$ of image patches. In this paper, we separate the real part and the imaginary part of the channel. The CSI image is rewritten as $\textbf{H}_\text{c} \in \mathbb{R}^{2 \times N_\text{c} \times N_\text{r}}$, where the first dimension in terms of 2 corresponds to the real part $\mathcal{R}(\textbf{H}_\text{c})$ and the imaginary part $\mathcal{I}(\textbf{H}_\text{c})$. To evaluate the probability of pixel values in $\textbf{H}_\text{c}$, we divide each of $\mathcal{R}(\textbf{H}_\text{c})$ and $\mathcal{I}(\textbf{H}_\text{c})$ into $(N_\text{c} - n +1) \times (N_\text{r} - n +1)$ patches. 

Without loss of generality, for the $j$th patch $\textbf{p}_j$, let $\mathcal{N}_j$ denote the set of neighboring patches of $\textbf{p}_j$, including $\textbf{p}_j$ itself. Besides, the Manhattan radius, denoted by $R$, is used to control the number of neighboring patches to determine the boundary of $\mathcal{N}_j$. Also, the $\mathcal{N}_j$ is a local region centered at $\textbf{p}_j$ and contains $(2R+1)^2$ patches. Let $\textbf{p}'_{j,r} \in \mathcal{N}_j$ denote the $r$th neighboring patch of $\textbf{p}_j$ for $r=\left \{1,2,\ldots ,(2R+1)^2 \right \} $.  To better elaborate the relationship between the Manhattan radius, $R$, and the boundary of $\mathcal{N}_j$, we exhibit the range of the neighboring patches set for a specific patch $\textbf{p}_j$ with $R=1$ and $R=3$ as shown in Fig. 2(a) and Fig. 2(b). It is observed that when the Manhattan radius $R=1$, the neighboring patches set contains $9$ patches, including $\textbf{p}_j$ itself. When the Manhattan radius $R=3$, the neighboring patches set contains totally $49$ patches.

To estimate $q_j$, we assume that all the $\textbf{p}'_{j,r}$, including $\textbf{p}_j$, obey the same distribution, i.e., $\textbf{p}'_{j,r} \sim q_j$. Then, we can adopt the Monte-Carlo method to estimate the probability $q_j$ as follows
\begin{equation}
    \widehat{q}_j = \frac{1}{(2R+1)^2} \sum_{{\textbf{p}'_{j,r}} \in \mathcal{N}_j} K(\textbf{p}_j,\textbf{p}'_{j,r}),
\end{equation}

\noindent where $K(\cdot,\cdot)$ is a kernel function. In this paper, we choose the Gaussian kernel function defined as
\begin{equation}
    K(\textbf{p}_j,\textbf{p}'_{j,r}) = \frac{1}{\sqrt{2\pi}h} \exp (- \left \| \textbf{p}_j-\textbf{p}'_{j,r}\right \|_2^2/2h^2),
\end{equation}

\noindent where $h$ denotes the bandwidth, controlling the radial range of an action. By exploiting the probability estimate $\widehat{q}_j$ in (6) and using (5), an estimate of the self-information $\widehat{I}_j$ is given by
\begin{equation}
    \widehat{I}_j = - \log_2 \frac{1}{(2R+1)^2}  \sum_{\textbf{p}'_{j,r} \in \mathcal{N}_j} \frac{1}{\sqrt{2\pi}h} \text{e}^{- \left \| \textbf{p}_j-\textbf{p}'_{j,r}\right \|_2^2/2h^2}  + \text{const}.
\end{equation}

\noindent By evaluating $ \widehat{I}_j$ for all the patches in $\mathcal{R}(\textbf{H}_\text{c})$ and $\mathcal{I}(\textbf{H}_\text{c})$, we can obtain a self-information matrix of $\textbf{H}_\text{c}$, denoted by $\textbf{I}(\textbf{H}_\text{c}) \in \mathbb{R}^{2 \times (N_\text{c}-n+1) \times (N_\text{r}-n+1)} $. Note that the self-information matrix reflects more directly the amount of information of each patch in the CSI image. By setting a self-information threshold $T$, it is possible to select the elements in $\textbf{I}(\textbf{H}_\text{c})$ with small self-information value and delete them, i.e., removing less informative entries in the CSI, for subsequent accurate compression, which will be elaborated with details in Section \uppercase\expandafter{\romannumeral4}.

\begin{figure*}[t]
\centering
\includegraphics[scale=0.32]{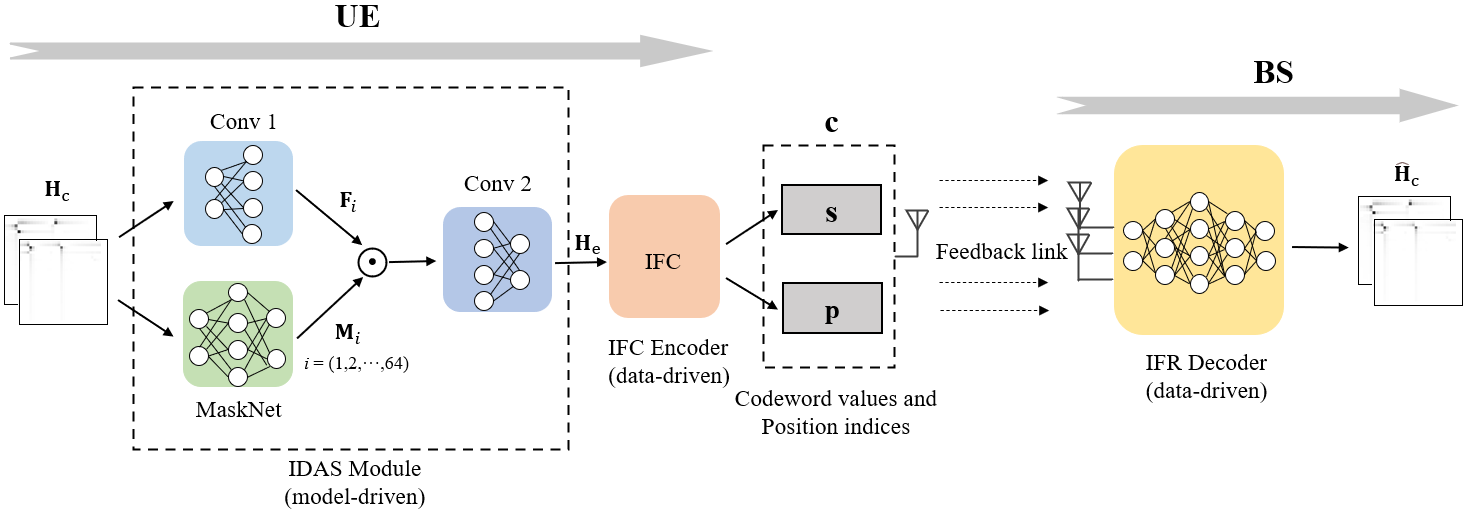}
\caption{The structure of the proposed IdasNet.}
\label{fig:label3}
\end{figure*}

To better understand the informative redundancy and the essential information embedded in the CSI image, we visualize the real part realization of a channel in the COST $2100$ channel model \cite{6393523} in Fig. 2(c). The real part $\mathcal{R}(\textbf{H}_\text{c})$ in Fig. 2(c) contains several clusters and each cluster contains both essential information and trivial information. The essential information corresponds to the resolvable path in MIMO systems, which dominates the details of $\mathcal{R}(\textbf{H}_\text{c})$. The trivial informative entries correspond to the lower-power propagation paths, which contains insignificant details of  $\mathcal{R}(\textbf{H}_\text{c})$, referred to as informative redundancy. In Fig. 2(d) we also visualize $\mathcal{R}(\textbf{H}_\text{c})$ with the informative redundancy removed based on the threshold. Each dotted grid in Fig. 2(d) is a patch $\textbf{p}_j$. 

In the IDAS module, we choose a value of the number  of texture patches contained in the CSI image as a prior to help determine which patches are the texture ones. Once the number of texture patches is set, the IDAS module selects the corresponding number of patches which are with the smallest self-information values as the texture patches. Then we choose the largest self-information value of the texture patches as the threshold, i.e., $T$, for effectively ruling out the informative redundancy. We can see that the informative redundancy of $\mathcal{R}(\textbf{H}_\text{c})$ is removed efficiently in Fig. 2(d) and the remaining essential information is obvious, which demonstrates its effectiveness to the subsequent accurate compression operation.

\begin{figure*}[t]
\centering
\includegraphics[scale=0.3]{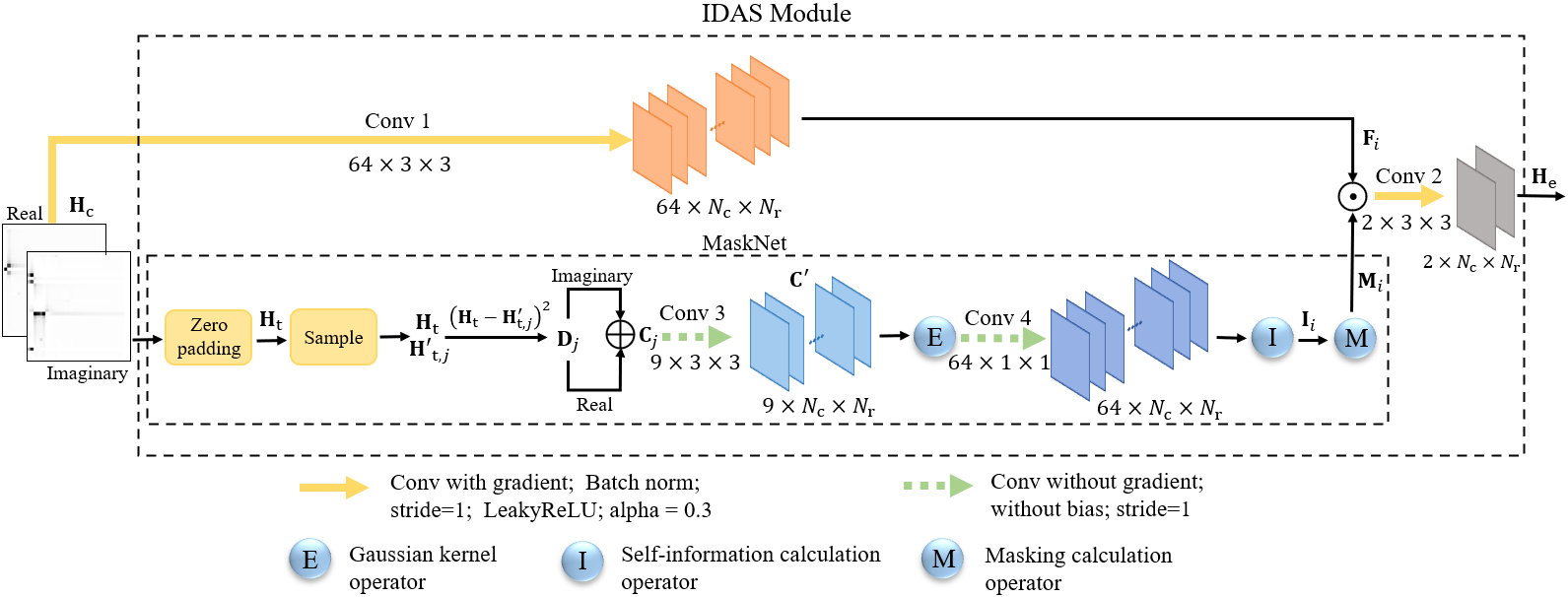}
\caption{The detailed structure of IDAS module. The operator “E” is based on (7). The operator``I'' is based on (6) and (8). The operator “M” is used to obtain masking matrices $\textbf{M}_i$.}
\label{fig:label4}
\end{figure*}

\section{Proposed Architecture of IdasNet}
In this section, we elaborate the proposed framework of IdasNet for CSI compression and feedback. The architecture of the proposed IdasNet is shown in Fig. 3, which consists of three modules, namely IDAS module, IFC encoder, and IFR decoder. To remove informative redundancy, the self-information model-based IDAS module pre-compresses the original CSI image,  $\textbf{H}_\text{c}$, based on the estimate of  self-information. Then an encoder named  IFC compresses the pre-compressed CSI image to a codeword $\textbf{c}$. At the BS, a decoder named IFR exploits the received $\textbf{c}$ to recover the CSI image. Detailed elaboration of each module in IdasNet is as follows.

\subsection{IDAS Module}
For the input of CSI image, the self-information model-based IDAS module is the first processor in the proposed IdasNet which removes the informative redundancy in $\textbf{H}_\text{c}$, regarded as a procedure of pre-compression. The design of the IDAS module is shown in Fig. 3. The IDAS module contains three components, i.e., convolutional layer 1 (Conv1), MaskNet, and convolutional layer 2 (Conv2). In particular, Conv1 transforms the CSI image $\textbf{H}_\text{c}$ to 64 feature maps, denoted by $\textbf{F}_i \in \mathbb{R}^{ N_\text{c} \times N_\text{r}}$ for $i=\left \{1,2, \ldots,64 \right \}$, where each feature map represents a specific feature of $\textbf{H}_\text{c}$. The MaskNet generates 64 masking matrices of size as $N_\text{c} \times N_\text{r}$, denoted by $\textbf{M}_i$ for  $i=\left \{1,2,\ldots, 64\right \}$, with binary elements of $0$ and $1$. To remove the informative redundancy of $\textbf{F}_i$ from the perspective of self-information, we let $\textbf{F}_i \odot \textbf{M}_i$, where $\odot$ denotes Hadamard product. Then, Conv2 restores the obtained 64 feature maps without informative redundancy to a $2$-dimensional self-information image, denoted by $\textbf{H}_\text{e} \in \mathbb{R}^{2 \times N_\text{c} \times N_\text{r}}$. 

\begin{algorithm}[t]
\caption{Algorithm of IFC Encoder }
\label{alg1}
\hspace*{0.02in} {\bf Input:}  The $2$-dimensional self-information image $\textbf{H}_\text{e}$. \\
\hspace*{0.02in} {\bf Output:} the codeword $\textbf{c}$ = [$\textbf{s} \ \textbf{p}$].  \\
\hspace*{0.02in} {\bf Parameters:} $M$, $v_i$, $p_i$, $s_i$.

\begin{algorithmic}[1]
\STATE Reshape $\textbf{H}_\text{e}$ to a vector $\textbf{v}$
\STATE Sort the elements $v_i$ in $\textbf{v}$ in descending order
\STATE Determine $M$ by (10)
    \FOR{$i=1, 2,\cdot \cdot \cdot, M$}
        \STATE Choose sorted $v_i$ as the $i$th codeword value $s_i$
        \STATE Store position index $p_i$ of $v_i$
    \ENDFOR
\end{algorithmic}
\end{algorithm}  %%%%%%%%%IFC encoder 算法表
In Fig. 4, we elaborate the design of three components in the IDAS module. Conv1 applies convolutional operations with a filter size of $64  \times 3 \times 3$ to yield the 64 feature maps. The parameters of Conv1 will be trained by using typical back propagation algorithms in the IdasNet, hence we refer to these parameters as network parameters with a gradient update in the rest of this paper. Note that zero padding is added before the convolution to ensure that the length and the width of the output tensor are the same as that of the input tensor. Moreover, Conv1 adopts batch normalization (BN) to stabilize and accelerate training. Considering that the output values are bipolar, it uses the LeakyReLU (LReLU) activation function rather than a simple ReLU function. The LReLU activation function serves as a nonlinear transformation in the network, which is defined as

\begin{eqnarray}
{\rm LeakyReLU}(x) = 
\left\{
\begin{array}{lll}
x,  \qquad x \geq 0 \\
0.3x, \ \ x \leq 0.
\end{array}
\right.
\end{eqnarray}

\begin{figure*}[!t]
\centering
\includegraphics[scale=0.30]{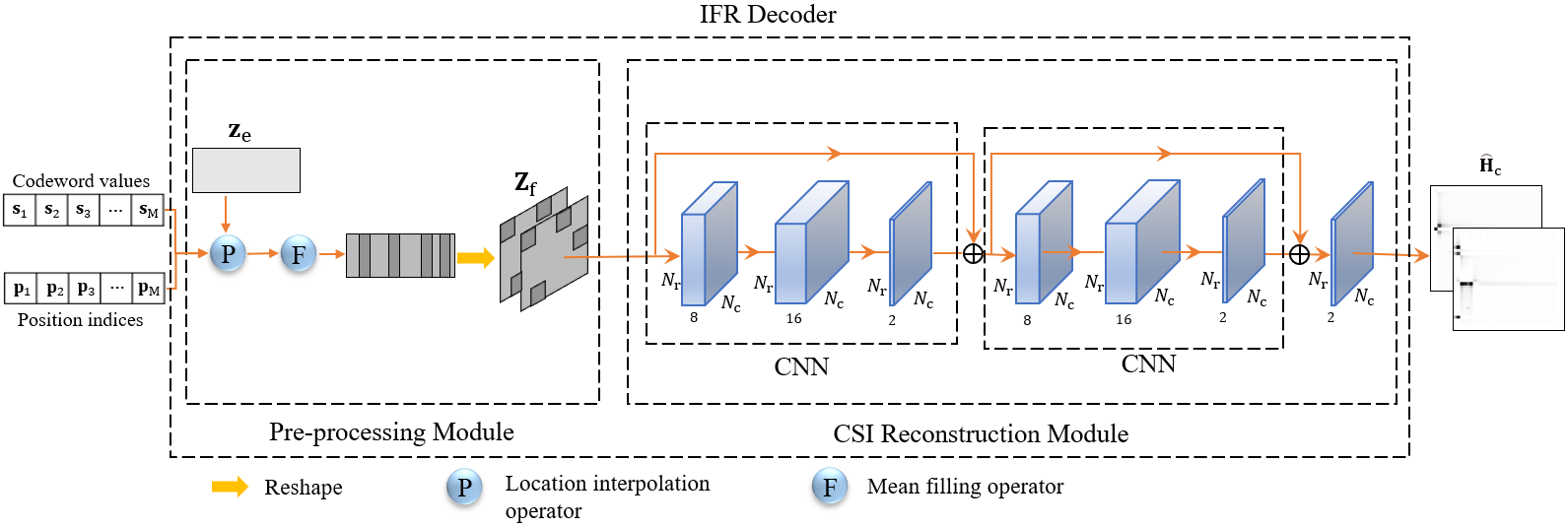}
\caption{The structure of IFR decoder. The operator ``P'' and ``F'' are based on (11). }
\label{fig:label5}
\end{figure*} %%%%%%%%IFR decoder详细结构图

On the other hand, the MaskNet consists of convolutional layer 3 (Conv3), operator ``E'', convolutional layer 4 (Conv4), operator ``I'', and operator ``M''. To simplify the required calculations, we regard a pixel in the CSI image $\textbf{H}_\text{c}$ as a patch. Then by applying zero padding we obtain an extended CSI image, denoted by $\textbf{H}_\text{t} \in \mathbb{R}^{2 \times (N_\text{c}+R-1) \times (N_\text{r}+R-1)}$. For each pixel in $\textbf{H}_\text{t}$, we randomly sample $9$ neighboring pixels based on the Manhattan radius $R$, and then the sampled CSI matrices, denoted by $\textbf{H}'_{\text{t},j} \in \mathbb{R}^{2 \times (N_\text{c}+R-1) \times (N_\text{r}+R-1)}$ for $j=\left \{1,2,\ldots,9 \right \}$, are formed by these neighboring pixels. The difference matrices, defined by $\textbf{D}_j \in \mathbb{R}^{2 \times (N_\text{c}+R-1) \times (N_\text{r}+R-1)} \triangleq (\textbf{H}_\text{t}-\textbf{H}'_{\text{t},j})^2$ for $j=\left \{1,2,\cdot \cdot \cdot,9 \right \}$, are then obtained by calculating the square of difference between $\textbf{H}_\text{t}$ and $\textbf{H}'_{\text{t},j}$. We then have $\textbf{C}_j \in \mathbb{R}^{(N_\text{c}+R-1) \times (N_\text{r}+R-1)} \triangleq \mathcal{R}(\textbf{D}_j) + \mathcal{I}(\textbf{D}_j)$ for $j=\left \{1,2,\cdot \cdot \cdot,9 \right \}$, by adding the real part information and the imaginary part information in terms of $\textbf{D}_j$.

The subsequent Conv3 is a mapping layer with a filter size of $9  \times 3 \times 3$, which is used to further map $\textbf{C}_j$ to $\textbf{C}' \in \mathbb{R}^{9 \times N_\text{c} \times N_\text{r}}$. Note that Conv3 is a mapping with fixed filter parameters. It is not involved for back propagation of the proposed IdasNet and it does not require gradient update and bias,  represented by dotted lines in Fig. 4. Then operator ``E'' calculates the corresponding Gaussian kernel matrices by using (7).  Conv4 with filter size of $64  \times 1 \times 1$ maps the Gaussian kernel matrices to 64 feature maps and it also has no need for gradient update and bias. Then the operator ``I''  calculates 64 self-information matrices, denoted by $\textbf{I}_i$ for $i=\left \{1,2,\ldots, 64\right \}$, based on the 64 feature maps by using (6) and (8). 

By setting a self-information threshold $T$, the operator ``M'' generates a masking matrix with size of $64 \times N_\text{c} \times N_\text{r}$. It forces the positions of elements in $\textbf{I}_i$ with the self-information smaller than $T$ to $0$, and sets the positions of other elements to $1$. Then we obtain the 64 masking matrices $\textbf{M}_i$ which contains only $0$ and $1$. The output of Conv1, $\textbf{F}_i$, is Hadamard producted by the corresponding masking matrix, $\textbf{M}_i$, to yield 64 new feature maps with the informative redundancy removed. Finally, Conv2 with filter size of $2  \times 3 \times 3$ restores the 64 feature maps without informative redundancy to a 2-dimensional self-information image $\textbf{H}_\text{e}$. Moreover, Conv2 also uses BN and LReLU activation function in (9).

\subsection{Encoder Design of IFC}
As shown in Fig. 3, the IFC encoder outputs the codeword $\textbf{c}$ = [$\textbf{s} \ \textbf{p}$]. The codeword values, $\textbf{s} \in \mathbb{R}^{M \times 1}$, consists of the selected elements with large self-information value in $\textbf{H}_\text{e}$, and the corresponding position indices, $\textbf{p} \in \mathbb{R}^{M \times 1}$, consists of the position in $\textbf{H}_\text{e}$ of each codeword value. The 2-dimensional self-information image $\textbf{H}_\text{e}$ deletes the elements with small self-information value and retains the other elements by utilizing the IDAS module. Hence, $\textbf{H}_\text{e}$ contains only the essential information of $\textbf{H}_\text{c}$. The calculation procedure of the IFC encoder is shown in Algorithm 1. First, $\textbf{H}_\text{e}$ is reshaped to a vector, denoted by $\textbf{v} \in \mathbb{R}^{2  N_\text{c}  N_\text{r} \times 1}$, and then the elements in  $\textbf{v}$ are arranged in a descending order based on the values of self-information. According to a predetermined compression ratio, the elements with larger self-information value are stored in  $\textbf{s}$ and their corresponding position indices are stored in $\textbf{p}$. The entire codeword $\textbf{c}$ is fed back to the IFR decoder for CSI reconstruction.

\begin{algorithm}[t]
\setstretch{1.0}
\caption{Algorithm of Pre-processing Module }
\label{alg3}
\hspace*{0.02in} {\bf Input:} The codeword values $\textbf{s}$ and position indices $\textbf{p}$.\\
\hspace*{0.02in} {\bf Output:} The 2-dimensional image $\textbf{Z}_\text{f}$.\\
\hspace*{0.02in} {\bf Parameters:} $\textbf{z}_\text{e}$, $z_i$, $y_i$, $s_i$, $p_i$, $\rho$.

\begin{algorithmic}[1]
\STATE Establish the all-zero vector $\textbf{z}_\text{e}$
\FOR{each element $z_i$ in $\textbf{z}_\text{e}$}
        \IF{$y_i$ = $p_i$ }
            \STATE $z_i$ = $s_i$
        \ELSE
            \STATE $z_i$ = $\rho$
        \ENDIF
\ENDFOR 
\STATE Reshape filled $\textbf{z}_\text{e}$ to 2-dimensional image $\textbf{Z}_\text{f}$
\end{algorithmic}
\end{algorithm}
%%%%%%%%%预填充模块算法表
Due to the fact that the IFC encoder feeds back not only $\textbf{s}$, but also its corresponding $\textbf{p}$, it is necessary to take $\textbf{p}$ into account when calculating the compression ratio. For fair comparison,  the compression ratio of the proposed IdasNet is calculated as
\begin{equation}
    \sigma = \frac{k_1 \times M + k_2 \times M} {k_1 \times 2 \times N_\text{c} \times N_\text{r}},
\end{equation}

\begin{table}[!t]  
\caption{The Parameters of CSI Reconstruction Module in Fig. 5}
\centering
\resizebox{75mm}{22mm}{
\begin{tabular}{crc}
\toprule
\multicolumn{3}{l}{{\bf Input}: The 2-dimensional image $\textbf{Z}_\text{f}$} \\  
\midrule
\multicolumn{3}{l}{\bf Convolutional}  \\ 
 {\bf Layers} & {\bf Filters/Stride/Padding} & {\bf Activation}\\
 1 & $8\times 3\times3/1/1$   & BN + $\text{LReLU}_{(0.3)}$   \\  
 2 & $16\times 3\times3/1/1$  & BN + $\text{LReLU}_{(0.3)}$   \\  
 3 & $2\times 3\times3/1/1$  & BN + $\text{LReLU}_{(0.3)}$   \\  
 4 & $8\times 3\times3/1/1$  & BN + $\text{LReLU}_{(0.3)}$   \\  
 5 & $16\times 3\times3/1/1$ & BN + $\text{LReLU}_{(0.3)}$   \\  
 6 & $2\times 3\times3/1/1$ & BN + $\text{LReLU}_{(0.3)}$   \\ 
 7 & $2\times 3\times3/1/1$ & BN + $\text{Sigmoid}$   \\ 
\midrule
\multicolumn{3}{l}{{\bf Output}: the CSI reconstruction image $\widehat{\textbf{H}}_\text{c}$}\\
\bottomrule  
\end{tabular}}
\label{tb:label1}
\end{table}

\noindent where $M$ denotes the number of codeword values in $\textbf{s}$, equivalently, the number of position indices. $k_1$ represents the number of bits to transmit each codeword value, and $k_2$ represents the number of bits to transmit each position index. 

\subsection{Decoder Design of IFR}
The IFR decoder is designed for reconstructing the CSI image, which is deployed at the BS. The IFR decoder consists of a pre-processing module and a CSI reconstruction module. The detailed structure of IFR decoder is shown in Fig. 5.

In the pre-processing module as described in Algorithm 2, we initialize $\textbf{z}_\text{e}$ as an all-zero vector with size of $2 N_\text{c} N_\text{r} \times 1$, which has the same dimension as $\textbf{v}$. By utilizing the codeword $\textbf{c}$ received from the IFC encoder, the operator ``P'' fills the codeword values $\textbf{s}$ into $\textbf{z}_\text{e}$ according to the corresponding position indices $\textbf{p}$. The remaining positions of $\textbf{z}_\text{e}$ are all filled with a mean value $\rho$ of the original CSI image $\textbf{H}_\text{c}$ by operator ``F'', where $\rho =\frac{1}{2 N_\text{c} N_\text{r}} \sum_{i=1}^{2 N_\text{c} N_\text{r}} h_{\text{c},i}$, $h_{\text{c},i}$ is the $i$th element of the original image $\textbf{H}_\text{c}$. Mathematically, the operations of ``P'' and ``F'' yield
\begin{eqnarray}
z_i = 
\left\{
\begin{array}{lll}
s_i, \ \ {\rm{if}} \ y_i = p_i \\
\rho, \ \ \rm{otherwise},
\end{array}
\right.
\end{eqnarray}

\noindent where $z_i$ represents the $i$th element of $\textbf{z}_\text{e}$ and $y_i$ represents the position index of $z_i$ for $i=\left \{1,2,\cdot \cdot \cdot, 2  N_\text{c}  N_\text{r} \right \}$, $s_i$ denotes the $i$th codeword value in $\textbf{s}$, and $p_i$ denotes the $i$th position index in $\textbf{p}$. The obtained $\textbf{z}_\text{e}$ in (11) is then reshaped to a 2-dimensional image, denoted by $\textbf{Z}_\text{f} \in \mathbb{R}^{2 \times N_\text{c} \times N_\text{r}}$, with the same size of $\textbf{H}_\text{c}$.

Following the structural design of the previous pre-processing module, the module of CSI reconstruction contains two consecutive components of convolutional neural networks (CNN) and ends with a layer of normalization. Each of the component of CNN consists of $3$ convolutional layers with filter sizes of $8  \times 3 \times 3$, $16  \times 3 \times 3$, and $2  \times 3 \times 3$, respectively, as shown in Table \uppercase\expandafter{\romannumeral1}. In order to prevent gradient vanishing during the training of CSI reconstruction, a shortcut connection is applied between the two components inspired by ResNet \cite{He_2016_CVPR}. Also, the normalization layer is used to scale the output of the second CNN into the range of $[0, 1]$. Mathematically, given an image $\textbf{A} \in \mathbb{R}^{C \times H \times W}$ and a kernel $\textbf{K} \in \mathbb{R}^{C \times H \times W}$, the 2D convolution $\textbf{A} \otimes \textbf{K}$ is defined as

\begin{figure*}[t]
\centering
\includegraphics[scale=0.31]{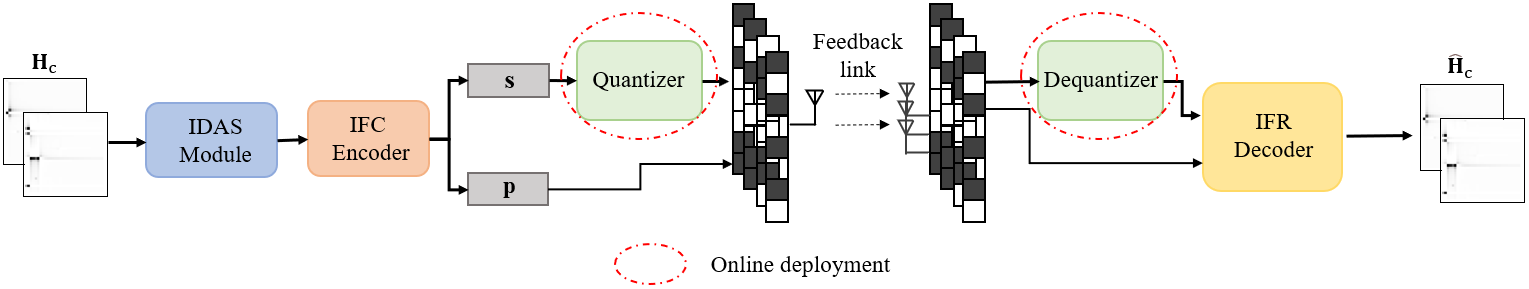}
\caption{IdasNet training with CSI quantization.}
\label{fig:label6}
\end{figure*}
%%%%%%%%量化反馈结构图
\begin{equation}
    (\textbf{A} \otimes \textbf{K})_{i,j} = \sum_{i_\text{h}=1}^H \sum_{i_\text{w}=1}^W \sum_{i_\text{c}=1}^C  \textbf{K}_{i_\text{c},i_\text{h},i_\text{w}} \textbf{I}_{i_\text{c},i+i_\text{h}-1,j+i_\text{w}-1}.
\end{equation}
As shown in Table \uppercase\expandafter{\romannumeral1}, we consecutively apply the BN and the LReLU activation function after each layer of convolutional operations. 

\subsection{Training}
We adopt a joint training of the IdasNet. For ease of elaboration, we denote the parameters of IdasNet as $\bf{\Phi}$ = $\lbrace \bm{\varphi}_{\text{IDAS}}, \bm{\varphi}_{\text{IFC}}, \bm{\varphi}_{\text{IFR}} \rbrace$, where $ \bm{\varphi}_{\text{IDAS}}$, $\bm{\varphi}_{\text{IFC}}$, and $\bm{\varphi}_{\text{IFR}}$ are the parameters of IDAS module, IFC encoder, and IFR decoder, respectively. The reconstructed CSI image is denoted by
\begin{equation}
    \widehat{\textbf{H}}_\text{c} = f(\textbf{H}_\text{c};\bm{\Phi})  \triangleq f_{\rm{IFR}}(f_{\text {IFC}}(f_{\text{IDAS}}(\textbf{H}_\text{c};\bm{\varphi}_{\text{IDAS}});\bm{\varphi}_{\text {IFC}});\bm{\varphi}_{\text{IFR}}),
\end{equation}

\noindent where $f_{\text{IDAS}}$ denotes the function of the IDAS module, $f_{\text{IFC}}$ denotes the function of the IFC encoder, and  $f_{\text{IFR}}$ denotes the function of the IFR decoder. Note that the input and output of IdasNet are normalized CSI image, whose elements are scaled in $[0,1]$. Besides, the $Adam$ optimizer \cite{kingma2017adam} is used to train IdasNet  and the loss function of mean squared error (MSE) is exploited for gradient update. The loss function is given as
\begin{equation}
    \text{Loss}(\bm{\Phi}) = \frac{1}{D} \sum_{i=1}^{D} \left \| f_{\text{IFR}}(f_{\text{IFC}}(f_{\text{IDAS}}(\textbf{H}_\text{c}[i]))) - \textbf{H}_\text{c}[i] \right \| _2^2,
\end{equation}

\noindent where $D$ denotes the total number of training samples in the training set. Finally, we exploit the NMSE to evaluate the performance of CSI reconstruction for IdasNet, which is defined as
\begin{equation}
    \text{NMSE} = E \left \{ \left \| \textbf{H}_\text{c} - \widehat{\textbf{H}}_\text{c} \right \|_2^2 / \left \| \textbf{H}_\text{c} \right \|_2^2 \right \}.
\end{equation}

\noindent Note that all the convolutional layers adopt kernels with a size of $3 \times 3$. In general, the kernel with size of $3 \times 3$ can extract information more accurately than kernels with sizes of $5 \times 5$ or $7 \times 7$ \cite{simonyan2015deep}. In particular, when calculating the self-information for the CSI image, both essential information and trivial information need to be considered. The kernels with sizes of $5 \times 5$ and $7 \times 7$ smoothen the trivial information when extracting feature and thus we choose the kernel with size of $3 \times 3$.

Due to the fact that transmitting continuous codeword values is difficult in practice, it is necessary to further quantize the continuous codeword values before performing feedback. For the proposed IdasNet, the output dimension of the IFC encoder and accordingly the input dimension of the IFR decoder are limited by the feedback channel capacity. For practical applications, it may be expected to train a common neural network such that it can adapt to a range of feedback rates. To achieve this goal, as shown in Fig. 6, we temporarily leave out the quantization operation during the offline training of IdasNet, which means that the codeword values are fed back to the IFR decoder without quantization. When the offline training is completed, we obtain the empirical probability distribution function (PDF) of the codeword values. In the stage of online deployment, we then exploit the Lloyd-Max algorithm \cite{1} to accomplish the quantization. At this stage, the codeword values are quantized through the quantizer and fed back to the IFR decoder. Finally, the IFR decoder at the BS recovers the CSI image by exploiting the quantized $\textbf{c}$ including quantized codeword values and corresponding position indices.

%% file: sections/04_experiments.tex
\section{Experimental Results}
\begin{figure*}[t]
\centering
\subfigure[$\sigma = \frac{1}{8}$.]{
\includegraphics[width = 7.2cm]{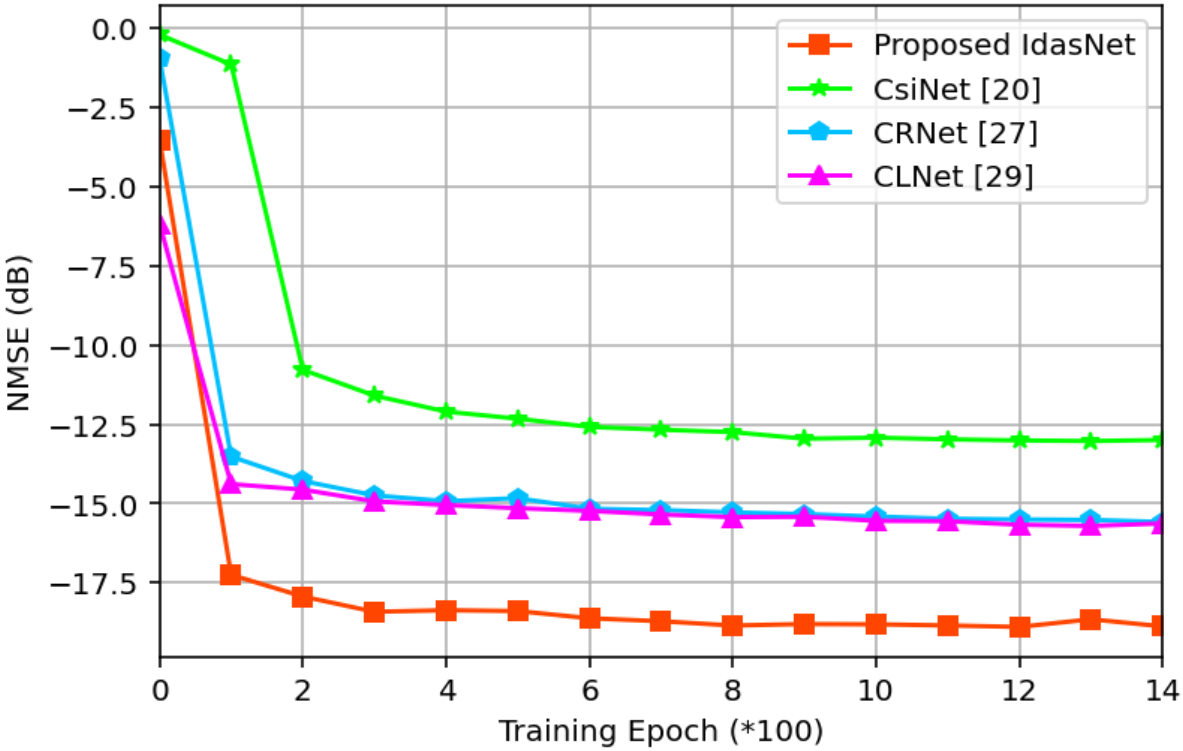}}
\subfigure[$\sigma = \frac{1}{16}$.]{
\includegraphics[width = 7cm]{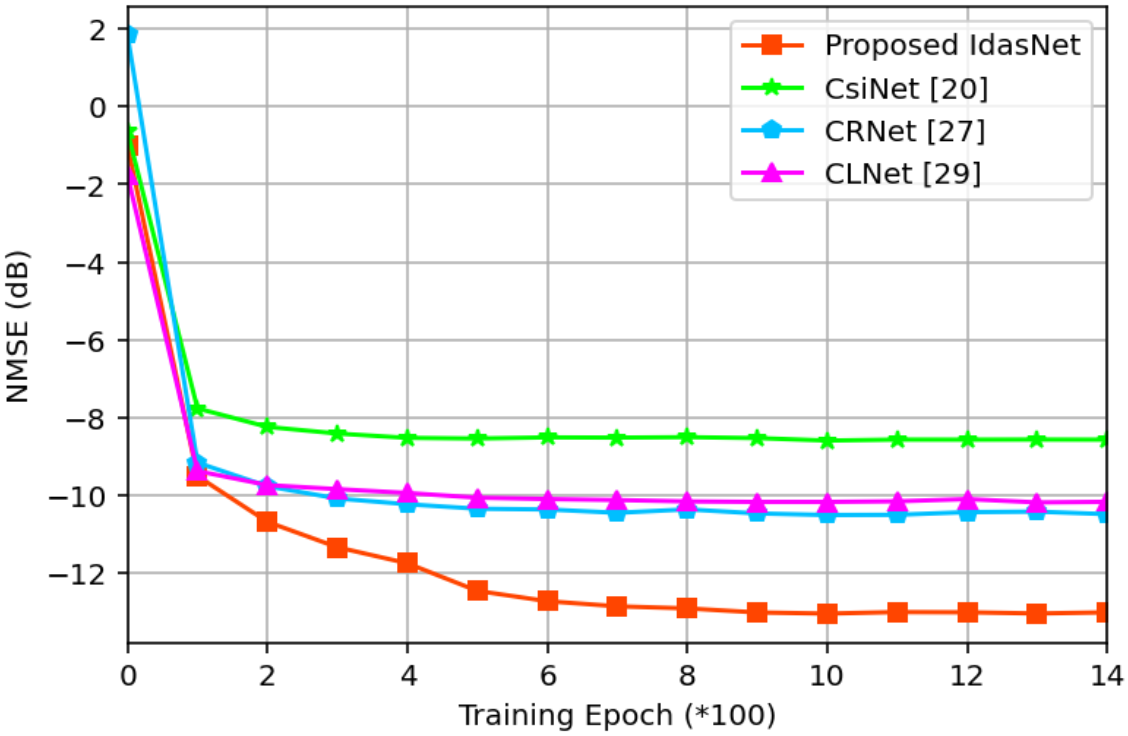}}

\subfigure[$\sigma = \frac{1}{32}$.]{
\includegraphics[width = 7.1cm]{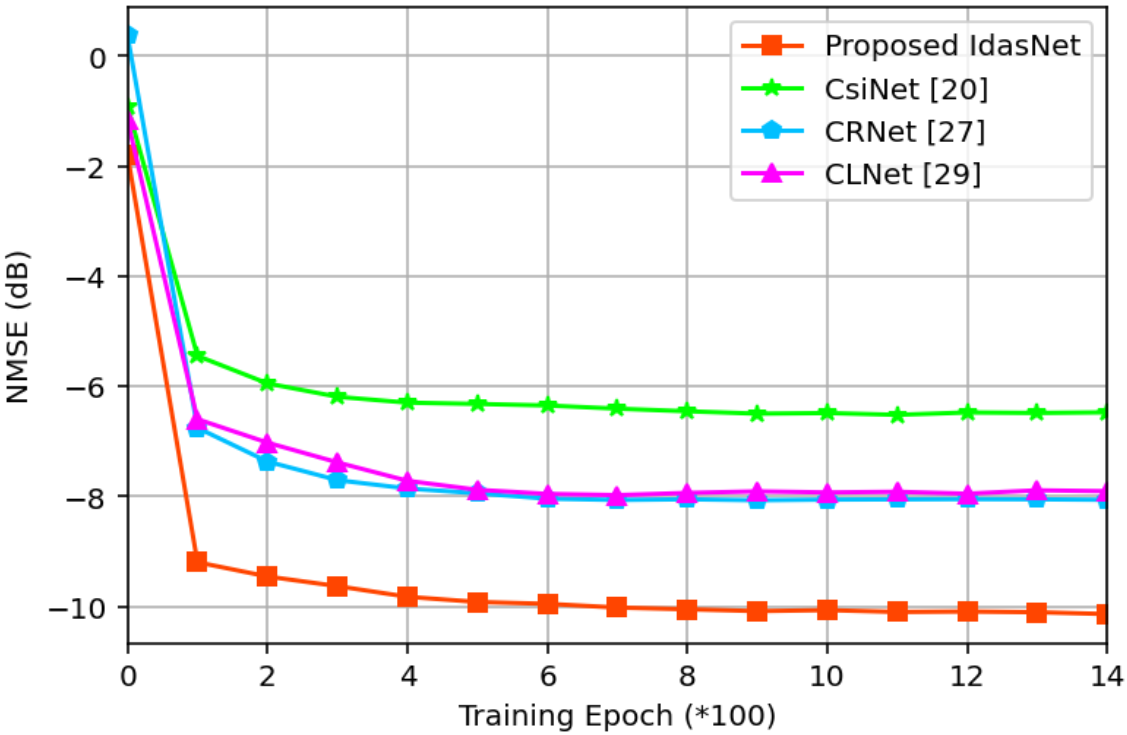}}
\subfigure[$\sigma = \frac{1}{64}$.]{
\includegraphics[width = 7cm]{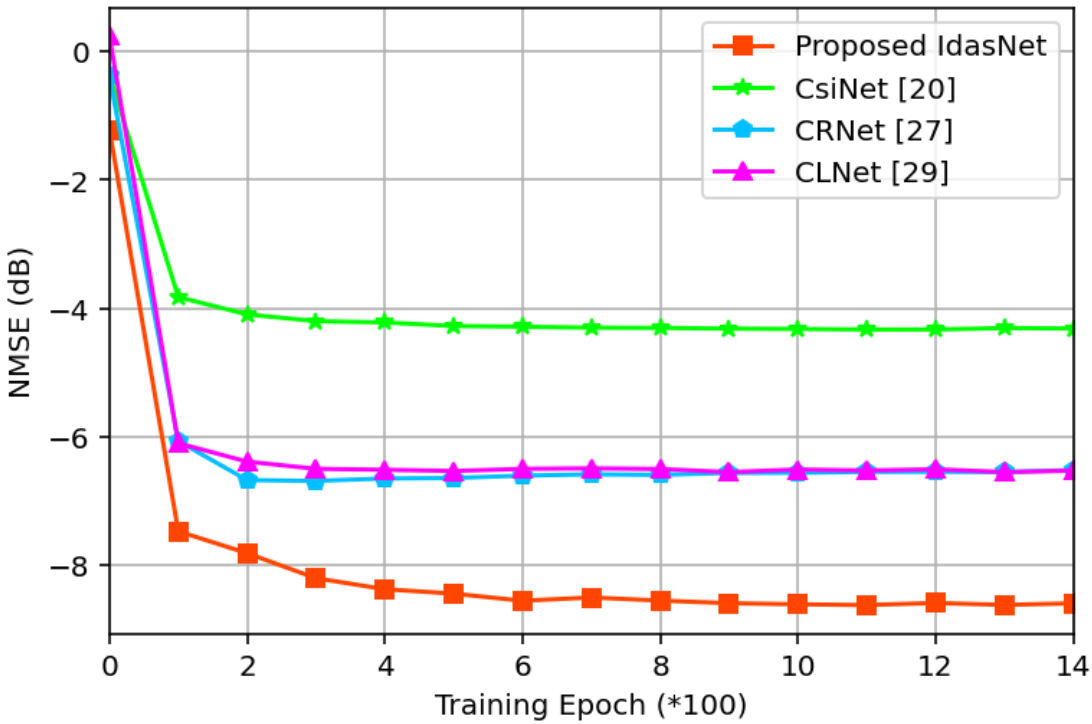}}
\caption{NMSE comparison under different values of compression ratio $\sigma$.}
\end{figure*}
In this section, we verify the effectiveness of the proposed IdasNet for CSI feedback. First, we describe the setting of parameters, the preparation of datasets and hardware facilities. Then, we present the performance comparison of the proposed IdasNet under different scenarios, which exhibits significantly better performance compared to existing DL methods. We also discuss the impacts of the number of feature maps in the IDAS module. In particular, we visualize the data distribution to verify the concept that the codeword acquired by the CSI image after removing informative redundancy is more conducive to reduce the error of the CSI reconstruction at BS. Also, we discuss the relation between the NMSE performance and the bit error rate (BER). Finally, we discuss the impact of different numbers of neighboring patches.

\subsection{Simulation Setup}
We generate the training set, validation set, and testing set through the COST $2100$ indoor channel model \cite{6393523}. The COST $2100$ channel model contains two environments: the indoor picocellular scenario at the $5.3$ GHz frequency band and the outdoor rural scenario at the $300$ MHz frequency band. The number of antennas at the BS is $N_\text{r} = 32$ and the number of subcarriers is $N_\text{s} = 1024$. When transforming the channel matrix into the angular-delay domain, we retain the first $32$ rows of the channel matrix, i.e., $N_\text{c}$ = $32$. The training set, validation set, and testing set contain respectively $100,000$, $30,000$, and $20,000$ samples.

\begin{table*}[!t]
\centering
\caption{Comparison of Transmitting Bits}
\resizebox{140mm}{23mm}{
\begin{tabular}{@{}c|c|cc|cc|c@{}}
\toprule
Methods &
  \begin{tabular}[c]{@{}c@{}}Compression\\ ratio\end{tabular} &
  \begin{tabular}[c]{@{}c@{}}Number of\\ codeword values\end{tabular} &
  \begin{tabular}[c]{@{}c@{}}Transmitting\\  bits\end{tabular} &
  \begin{tabular}[c]{@{}c@{}}Number of \\ position index\end{tabular} &
  \begin{tabular}[c]{@{}c@{}}Transmitting\\  bits\end{tabular} &
  \begin{tabular}[c]{@{}c@{}}Total bits\end{tabular} \\ \midrule
CLNet {[}29{]} & \multirow{2}{*}{1/8}  & 256          & 64 & 0            & 0  & 16,384 \\
\textbf{IdasNet}            &                       & 221+\textbf{1} & 64 & 221 & 10 & \textbf{16,418} \\ \midrule
CLNet {[}29{]} & \multirow{2}{*}{1/16} & 128          & 64 & 0            & 0  & 8,192  \\
\textbf{IdasNet}           &                       & 111+\textbf{1} & 64 & 111 & 10 & \textbf{8,278}  \\ \midrule
CLNet {[}29{]} & \multirow{2}{*}{1/32} & 64           & 64 & 0            & 0  & 4,096  \\
\textbf{IdasNet}           &                       & 56+\textbf{1}  & 64 & 56  & 10 & \textbf{4,208}  \\ \midrule
CLNet {[}29{]} & \multirow{2}{*}{1/64} & 32           & 64 & 0            & 0  & 2,048  \\
\textbf{IdasNet}            &                       & 28+\textbf{1}  & 64 & 28  & 10 & \textbf{2,136}  \\ \bottomrule
\end{tabular}}
\end{table*}

As for the design of IdasNet, we extract $64$ feature maps for self-information deletion and selection, which corresponds to the kernel with filter size of $64 \times 3 \times 3$ in Conv1 as shown in Fig. 4. When calculating the probability, we set the Manhattan radius as $R=3$. We set that the number of texture patches contained in each of $\mathcal{R}(\textbf{H}_\text{c})$ and $\mathcal{I}(\textbf{H}_\text{c})$ is $224$. All the testing samples are excluded from the training samples and validation samples. The trainable weights and bias of all the convolutional layers are initialized randomly, and the non-trainable weights are initialized to unit matrix. The $Adam$ optimizer is used. The number of epoch is set to $1400$, the batch size is set to $200$. The simulation is carried out in Pytorch on a GTX3090 GPU. 

Especially, the learning rate (lr) plays an important role for the convergence result of the network. In order to make the proposed network to learn the global optimal solution, the lr is linearly increased from zero to its maximal rate, which is called ``warm up" \cite{9149229}. After that, the lr descends like the cosine trend following as
\begin{equation}
    \gamma = \gamma_\text{min} + \frac{1}{2}(\gamma_\text{max} - \gamma_\text{min})(1 + \text{cos}(\frac{t-T_\text{w}}{T_\text{t}-T_\text{w}}\pi)),
\end{equation}
\noindent where $t$ denotes the index of current epoch. $\gamma_\text{min}$, $\gamma_\text{max}$, and $\gamma$ denote the initial, final, and current lr. $T_\text{t}$ and $T_\text{w}$ denote the numbers of total and warm up epochs, respectively.

\subsection{Performance Comparison}
To validate the performance of the proposed IdasNet for CSI feedback, we compare its performance with existing CSI feedback methods using DL. For comparison, we carry out the experimental simulations with various compression ratios of $1/8$, $1/16$, $1/32$, and $1/64$. We compare IdasNet with some existing DL methods for the CSI compression in terms of NMSE performance and the network complexity.

\begin{figure*}[t]
\centering
\includegraphics[scale=0.2]{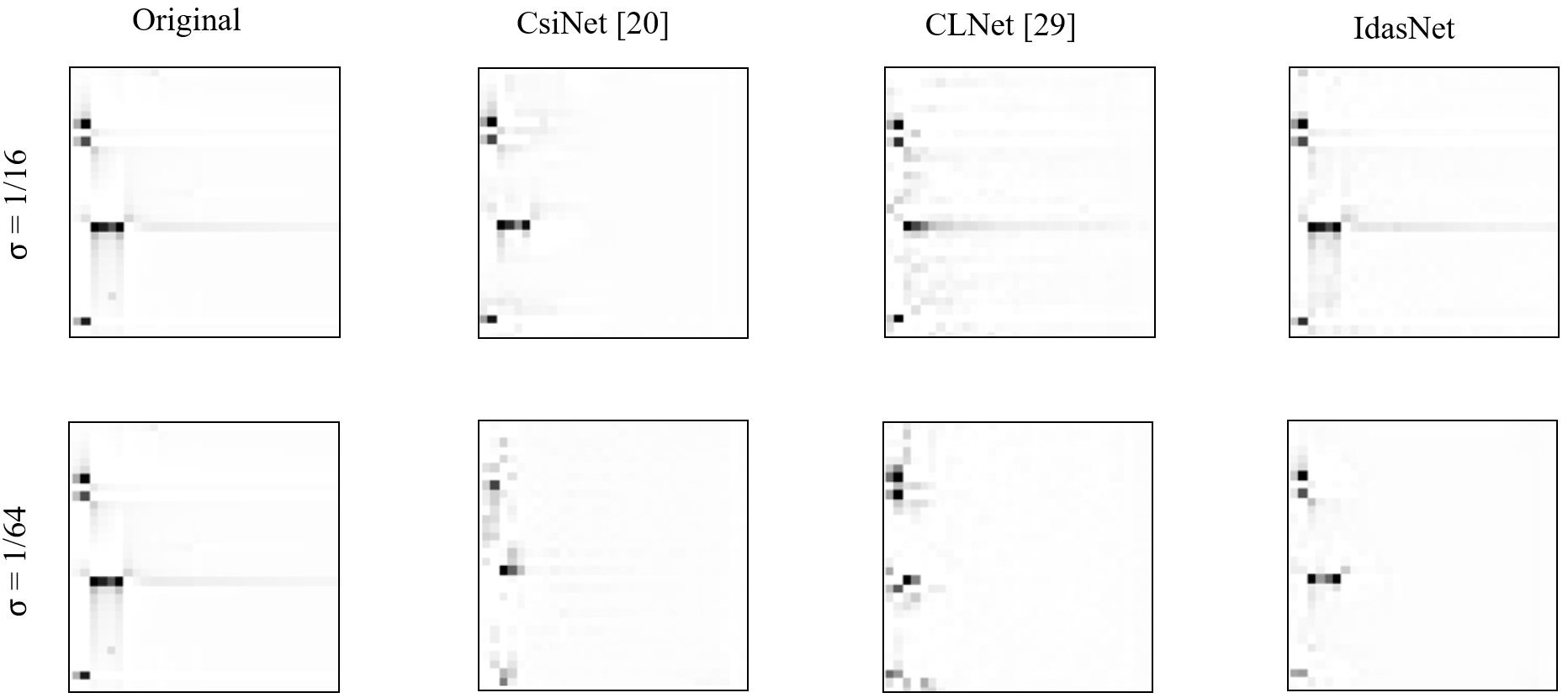}
\caption{Comparison with CSI reconstruction image under different compression ratios. }
\label{fig:label8}
\end{figure*}
%%%%%%%%复原CSI比较

\subsubsection{NMSE Performance}
To verify the effectiveness of the IdasNet on the CSI reconstruction, we compare the NMSE of the IdasNet with existing methods, including CsiNet \cite{8322184}, CRNet \cite{9149229}, and CLNet \cite{9497358}. The calculation of NMSE is based on (15). Comparison results are shown in Fig. 7. We observe that IdasNet outperforms CsiNet, CRNet, and CLNet under all the different compression ratios. Note that when the compression ratio $\sigma = 1/8$, the NMSEs of IdasNet and CLNet are respectively $-18.87$ dB and $-15.63$ dB, and the NMSE performance gain by IdasNet is approximately $3$ dB. When the compression ratio decreases to $\sigma = 1/16$, the NMSEs of IdasNet and CLNet are respectively $-13.51$ dB and $-10.17$ dB, and the performance gain of IdasNet is still $3$ dB. When the compression ratio further decreases to $\sigma = 1/64$, the performance gain of the proposed IdasNet remains to be $3$ dB in terms of the NMSE. As the compression ratio decreases, the interference of informative redundancy for CSI reconstruction is magnified, thus removing informative redundancy effectively is necessary for improving the accuracy of CSI recovery. This illustrates why the proposed IdasNet can still achieve high NMSE performance gain under the low compression ratio. Moreover, both the removement of informative redundancy and the replacement of a mean value, $\rho$, help improve the accuracy of CSI reconstruction.

Note that the proposed IFC encoder feeds back not only codeword values \textbf{s} to the IFR decoder, but also the corresponding position indices \textbf{p}. For a fair comparison, from the perspective of feedback signaling overhead, the position indices are considered in the calculation of the compression ratio. Different from the previous methods, the calculation of compression ratio of IdasNet is based on (10). The comparison of the number of the transmitting bits under different compression ratios is summarized in Table \uppercase\expandafter{\romannumeral2}. Each codeword value needs $64$ bits to transmit. As the number of positions with non-zero values in vector $\textbf{v}$ is less than $1024$, so the position index of each codeword value is limited to $10$ bits to transmit. Note that $\rho$, i.e., mean value of the original CSI image $\textbf{H}_\text{c}$, should be counted towards the codeword length as the feedback information, which corresponds to ``1'' in the number of codeword values in Table \uppercase\expandafter{\romannumeral2}. Also, $\rho$ is a double-floating number and needs at most 64 bits for representation, and there is no need for transmitting the position index of $\rho$.

We observe from Table \uppercase\expandafter{\romannumeral2} that when compression ratio $\sigma = 1/8$, the existing DL networks such as CLNet \cite{9497358} contain $256$ codeword values, thus the total transmitting bits amount to $16,384$ bits for CSI feedback. The IdasNet contains $222$ codeword values and $221$ position indices, and the total transmitting bits are $16,418$ bits, which is almost equal to that of the CLNet for a fair comparison. When compression ratio $\sigma  = 1/16$, the existing DL networks contain $128$ codeword values and the total transmitting bits are $8,192$ bits. The IdasNet contains $112$ codeword values and $111$ position indices, implying the total number of transmitting bits as $8,278$ bits. Similar setup of transmitting bits is configured as in Table \uppercase\expandafter{\romannumeral2} for $\sigma  = 1/32$ and $\sigma  = 1/64$. Although the codeword of IdasNet contains codeword values and position indices, the overhead of IdasNet is set the same as the comparing methods under all compression ratios. 

\begin{table*}[t]
\centering
\caption{Network Complexity Comparison}
% \resizebox{\textwidth}{15mm}{
% \setlength{\tabcolsep}{0.01mm}{
\resizebox{165mm}{18mm}{
\begin{tabular}{@{}c|ccc|ccc|ccc|ccc@{}}
\toprule
\begin{tabular}[c]{@{}c@{}}Compression\\ ratio\end{tabular} &
  \multicolumn{3}{c|}{$\sigma$=1/8} &
  \multicolumn{3}{c|}{$\sigma$=1/16} &
  \multicolumn{3}{c|}{$\sigma$=1/32} &
  \multicolumn{3}{c}{$\sigma$=1/64} \\ \midrule
\diagbox [trim=l] {Method} {Parameters} &
  \begin{tabular}[c]{@{}c@{}}Trainable\end{tabular} &
  \begin{tabular}[c]{@{}c@{}}Non-trainble\end{tabular} &
  \begin{tabular}[c]{@{}c@{}}Total\end{tabular} &
  \begin{tabular}[c]{@{}c@{}}Trainable\end{tabular} &
  \begin{tabular}[c]{@{}c@{}}Non-trainble\end{tabular} &
  \begin{tabular}[c]{@{}c@{}}Total\end{tabular} &
  \begin{tabular}[c]{@{}c@{}}Trainable\end{tabular} &
  \begin{tabular}[c]{@{}c@{}}Non-trainble\end{tabular} &
  \begin{tabular}[c]{@{}c@{}}Total\end{tabular} &
  \begin{tabular}[c]{@{}c@{}}Trainable\end{tabular} &
  \begin{tabular}[c]{@{}c@{}}Non-trainble\end{tabular} &
  \begin{tabular}[c]{@{}c@{}}Total\end{tabular} \\ \midrule
CsiNet [20] &
  1,052,626 &
  0 &
  \textbf{1,052,626} &
  528,210 &
  0 &
  \textbf{528,210} &
  266,002 &
  0 &
  \textbf{266,002} &
  134,898 &
  0 &
  \textbf{134,898} \\ \midrule
CRNet [27] &
  { 1,054,006} &
  0 &
  \textbf{1,054,006 } &
  { 529,590} &
  0 &
  \textbf{529,590 } &
  { 267,382} &
  0 &
  \textbf{267,382 } &
  { 136,278} &
  0 &
  \textbf{136,278 } \\ \midrule
CLNet [29] &
  2,105,538 &
  0 &
  \textbf{2,105,538} &
  1,056,578 &
  0 &
  \textbf{1,056,578} &
  532,162 &
  0 &
  \textbf{532,162} &
  269,954 &
  0 &
  \textbf{269,954} \\ \midrule
\textbf{IdasNet} &
  4202 &
  657 &
  \textbf{4859} &
  4202 &
  657 &
  \textbf{4859} &
  4202 &
  657 &
  \textbf{4859} &
  4202 &
  657 &
  \textbf{4859} \\ \bottomrule
\end{tabular}}
\end{table*}

\begin{figure*}[t]
\centering
\subfigure[$\sigma = \frac{1}{8}$.]{
\includegraphics[width = 7.2cm]{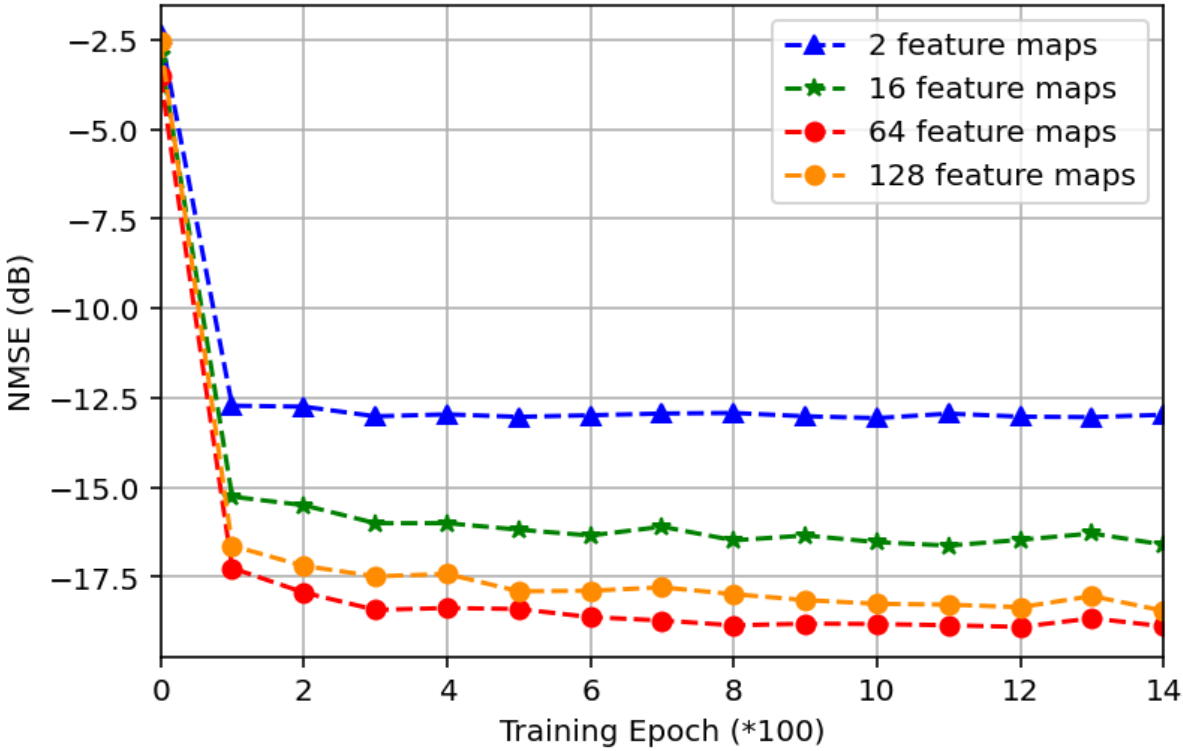}}
\subfigure[$\sigma = \frac{1}{16}$.]{
\includegraphics[width = 7cm]{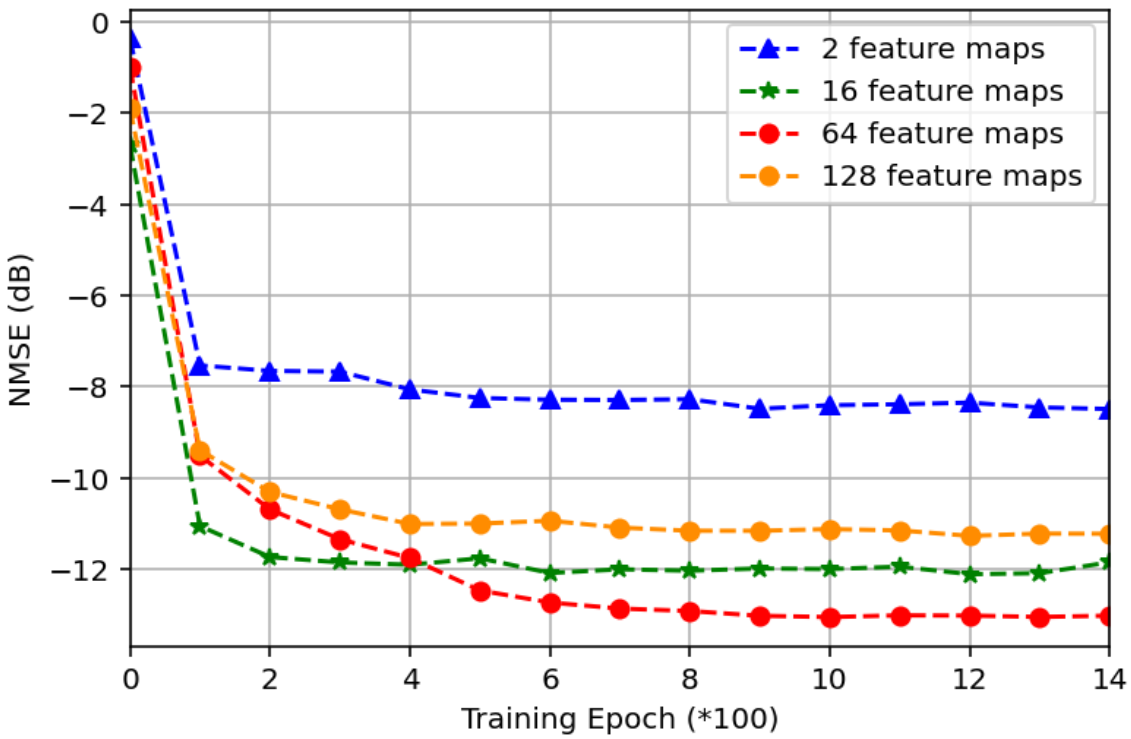}}

\subfigure[$\sigma = \frac{1}{32}$.]{
\includegraphics[width = 7.1cm]{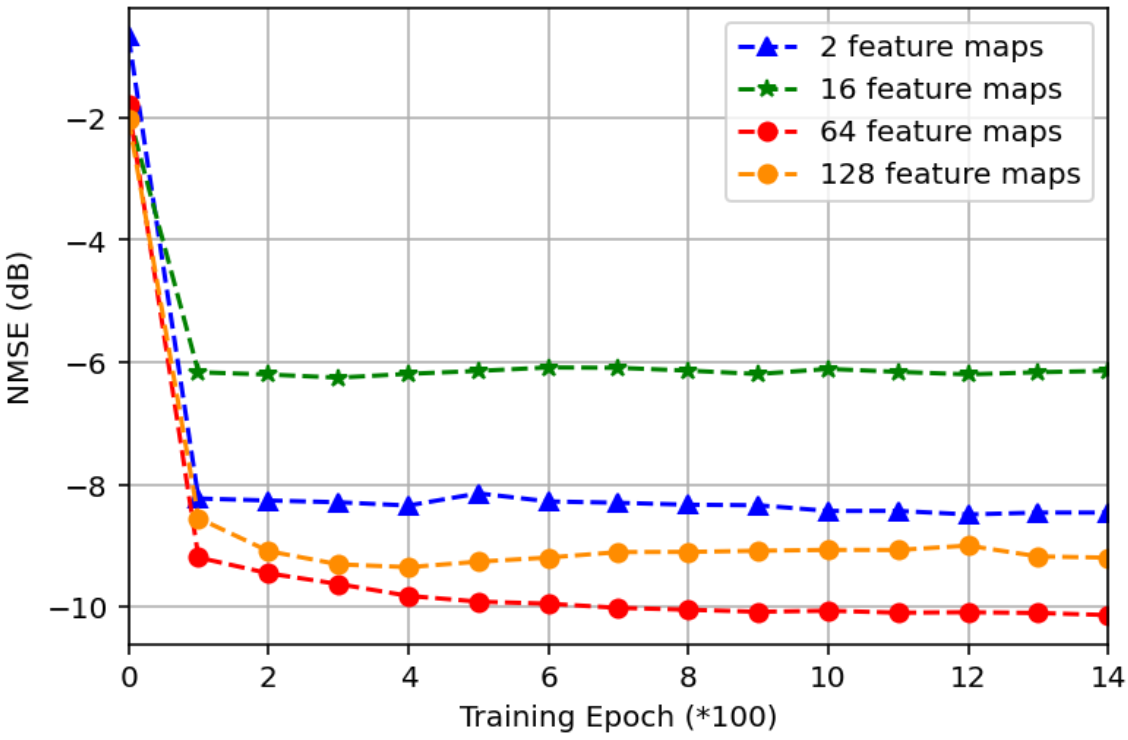}}
\subfigure[$\sigma = \frac{1}{64}$.]{
\includegraphics[width = 7cm]{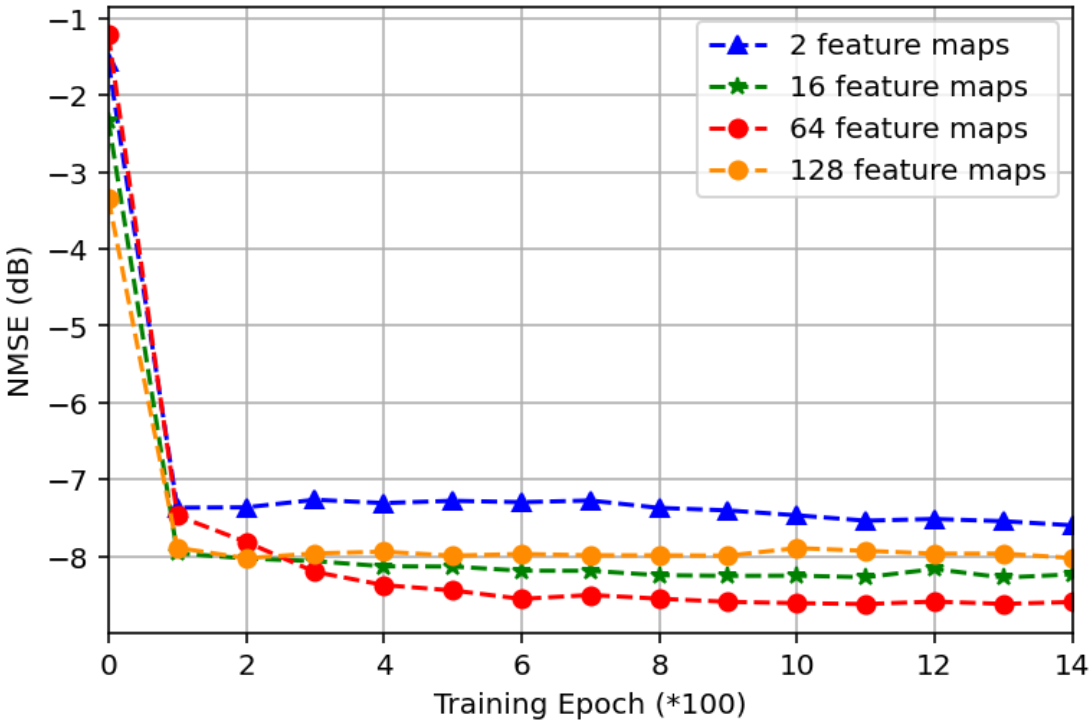}}

\caption{NMSE comparison under different numbers of feature maps.}
\label{fig:label9}
\end{figure*}

\subsubsection{Visualization of CSI Reconstruction}
In order to intuitively investigate the performance of CSI reconstruction of IdasNet, we visualize the reconstructed real and imaginary part by different methods in Fig. 8 under compression ratios of $1/16$ and $1/64$, where the strength of a pixel represents the magnitude of the channel gain. From Fig. 8 we observe that the IdasNet recovers the CSI image more accurately than that of CsiNet and CLNet. As the compression ratio decreases, better performance of the CSI reconstruction is achieved by IdasNet compared to both the existing methods.

\begin{table}[t]
\centering
\caption{NMSE Comparison With And Without Quantization Under Different Scenarios}
\resizebox{75mm}{36mm}{
\begin{tabular}{@{}cccc@{}}
\toprule
\quad Methods                        & \begin{tabular}[c]{@{}c@{}} $\sigma$ \end{tabular} & \begin{tabular}[c]{@{}c@{}} NMSE (dB)\\ Indoor/Outdoor \end{tabular}                            & \begin{tabular}[c]{@{}c@{}} NMSE-Q (dB)\\ Indoor/Outdoor \end{tabular}      \\ \midrule
\multicolumn{1}{c|}{CsiNet [20] } & \multicolumn{1}{c|}{\multirow{4}{*}{1/8}}                  & \multicolumn{1}{c|}{-12.99/-7.67}           & -12.73/-7.42           \\
\multicolumn{1}{c|}{CRNet [27]}  & \multicolumn{1}{c|}{}                                      & \multicolumn{1}{c|}{-15.58/-7.73}          & -15.36/-7.59          \\
\multicolumn{1}{c|}{CLNet [29] }    & \multicolumn{1}{c|}{}                                      & \multicolumn{1}{c|}{-15.63/-8.15}          & -15.41/-7.89          \\
\multicolumn{1}{c|}{\textbf{IdasNet}}   & \multicolumn{1}{c|}{}                                      & \multicolumn{1}{c|}{\textbf{-18.87/-10.34}} & \textbf{-18.62/-10.19} \\ \midrule
\multicolumn{1}{c|}{CsiNet [20] } & \multicolumn{1}{c|}{\multirow{4}{*}{1/16}}                 & \multicolumn{1}{c|}{-8.57/-4.32}           & -8.31/-4.14           \\
\multicolumn{1}{c|}{CRNet [27] }  & \multicolumn{1}{c|}{}                                      & \multicolumn{1}{c|}{-10.49/-5.37}           & -10.27/-5.13           \\
\multicolumn{1}{c|}{CLNet [29] }    & \multicolumn{1}{c|}{}                                      & \multicolumn{1}{c|}{-10.17/-5.46}           & -9.92/-5.21           \\
\multicolumn{1}{c|}{\textbf{IdasNet}}   & \multicolumn{1}{c|}{}                                      & \multicolumn{1}{c|}{\textbf{-13.51/-6.15}} & \textbf{-13.37/-5.87} \\ \midrule
\multicolumn{1}{c|}{CsiNet [20] } & \multicolumn{1}{c|}{\multirow{4}{*}{1/32}}                 & \multicolumn{1}{c|}{-6.47/-2.53}           & -6.25/-2.32           \\
\multicolumn{1}{c|}{CRNet [27] }  & \multicolumn{1}{c|}{}                                      & \multicolumn{1}{c|}{-8.06/-3.51}           & -7.88/-3.28           \\
\multicolumn{1}{c|}{CLNet [29] }    & \multicolumn{1}{c|}{}                                      & \multicolumn{1}{c|}{-7.90/-3.53}           & -7.73/-3.38           \\
\multicolumn{1}{c|}{\textbf{IdasNet}}   & \multicolumn{1}{c|}{}                                      & \multicolumn{1}{c|}{\textbf{-10.13/-5.03}}  & \textbf{-9.94/-4.91}  \\ \midrule
\multicolumn{1}{c|}{CsiNet [20]} & \multicolumn{1}{c|}{\multirow{4}{*}{1/64}}                 & \multicolumn{1}{c|}{-4.31/-1.96}           & -4.14/-1.72           \\
\multicolumn{1}{c|}{CRNet [27] }  & \multicolumn{1}{c|}{}                                      & \multicolumn{1}{c|}{-6.51/-2.15}           & -6.32/-1.84           \\
\multicolumn{1}{c|}{CLNet [29] }    & \multicolumn{1}{c|}{}                                      & \multicolumn{1}{c|}{-6.52/-2.17}           & -6.34/-1.87           \\
\multicolumn{1}{c|}{\textbf{IdasNet}}   & \multicolumn{1}{c|}{}                                      & \multicolumn{1}{c|}{\textbf{-9.34/-3.63}}  & \textbf{-9.18/-3.42}  \\ \bottomrule
\end{tabular}}
\end{table}

\subsubsection{Network Complexity}
We compare the network complexity in terms of the number of parameters. The comparison result is shown in Table \uppercase\expandafter{\romannumeral3}. We observe that IdasNet has far less number of parameters than the existing methods, including CsiNet, CRNet, and CLNet. The number of parameters in IdasNet is $3$ orders of magnitude smaller than that of CsiNet, CRNet and CLNet. The reason is that in these existing DL-based networks, both the encoder and the decoder adopt the fully-connected (FC) layer, which is the main contribution to the large number of parameters. Meanwhile, as the compression ratio varies, the output dimension of the FC layer also varies, resulting in the changes of the number of network parameters.

% As the compression ratio increases in CsiNet, CRNet, and CLNet, the number of model parameters increases rapidly.

In contrast, we design the IFC encoder and the IFR decoder to obtain compressed codeword and recover the CSI image without using FC layer, and the IdasNet connects an informative model-driven module, IDAS module, before a data-driven network for CSI compression and feedback. Hence, the total number of parameters of IdasNet is much less than the existing DL-based methods. Moreover, since the IFC encoder and IFR decoder do not adopt FC layer, so the total number of parameters of IdasNet is not affected by the compression ratio. In addition, there are two convolutional layers without gradient update in the IDAS module, which corresponds to non-trainable parameters in IdasNet.

\subsection{Impact of Feature Maps Choices}
From Section \uppercase\expandafter{\romannumeral3} we can observe that the self-information deletion and selection is applied for $64$ feature maps $\textbf{F}_i$ rather than the original CSI image $\textbf{H}_\text{c}$. It is intuitively that the number of feature maps affects the performance and training complexity of the network. It is essential for IdasNet to choose a proper number of feature maps. We evaluate the impact of the number of feature maps on the NMSE performance of IdasNet. The number of feature maps corresponds to the discussion of Conv1 as shown in Fig. 4.

We select $2$, $16$, $64$, and $128$ feature maps to evaluate the effect on the NMSE performance of IdasNet. The simulation results are shown in Fig. 9. We use the training dataset to train the network and use the testing dataset to evaluate the NMSE performance of different numbers of feature maps, which shares the same configuration as that used in Fig. 7. We observe that $64$ feature maps is optimal among all these selections for the NMSE performance of IdasNet. In fact, fewer feature maps cannot extract enough features from the original CSI image $\textbf{H}_\text{c}$, which results in that informative redundancy could not be precisely removed, and thus degrades the performance of IdasNet. In contrast, more feature maps than $64$ can result in extracting excessive feature from $\textbf{H}_\text{c}$. When the number of feature maps is so large that a part of essential information can be removed when processing their self-information. Therefore, utilizing too many feature maps can result in not only long training time for the network but also poor NMSE performance. From Fig. 9 we can see that choosing $64$ feature maps is a perfect choice under all compression ratios with CSI feedback.

\begin{figure}[t]
\centering
\subfigure[Input of existing encoder.]{
\includegraphics[width = 0.3\linewidth]{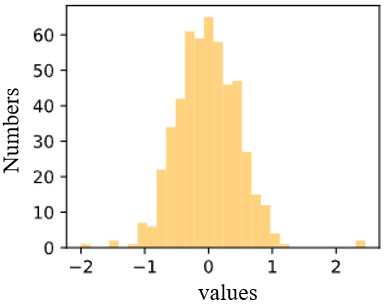}}
\subfigure[Input of proposed IFC encoder.]{
\includegraphics[width = 0.3\linewidth]{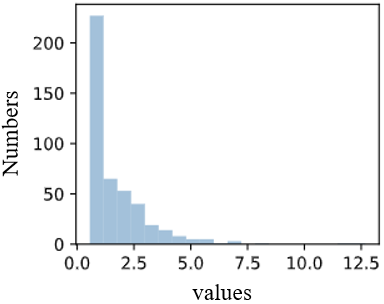}}

\caption{Data distribution comparison. }
\label{fig:label10}
\end{figure}

\subsection{Performance with Quantization Feedback}
For practical systems with limited feedback bandwidth, it is less possible to feed back continuous values of the codeword. In general, the codeword values should be quantized before sending back to the BS. For the offline training, we train the IdasNet without considering quantization. For the online deployment, we apply the Lloyd-Max algorithm to quantize the codeword values $\textbf{s}$. The comparison of NMSE with and without quantization is shown in Table \uppercase\expandafter{\romannumeral4}. NMSE-Q denotes the NMSE value with codeword values quantization. To achieve fair comparison, the CsiNet, the CRNet, and the CLNet are tested with quantization in the same way. From Table \uppercase\expandafter{\romannumeral4}, it is observed that the proposed IdasNet always outperforms CsiNet, CRNet, and CLNet even with quantization in the tests. We also observe that the difference between the NMSE with quantization and the NMSE without  quantization for each of the method is marginal with a reasonable value of feedback bits.

By comparing the NMSE performance of indoor scenario dataset and outdoor scenario dataset, we observe that the sparseness of the input CSI image has an impact on the efficiency for the proposed IdasNet and other existing DL-based methods. As discussed in CsiNet \cite{8322184}, the dataset of outdoor scenario is indeed more dense than the CSI dataset of indoor scenario. For a horizontal comparison in Table \uppercase\expandafter{\romannumeral4}, we observe that the NMSE adopting the indoor scenario dataset is significantly lower than the NMSE by using the outdoor scenario dataset for the same DL-based method. In fact, for DL-based CSI compression and feedback, the sparser channel is helpful for the encoder to better compress the CSI image and the BS can also recover the CSI image more easily. This validates the above statement that the sparseness of the input CSI image has an impact on the efficiency for DL-based methods. On the other hand, the NMSE performance of the proposed IdasNet still always outperforms the other methods including CsiNet \cite{8322184}, CRNet \cite{9149229}, and CLNet \cite{9497358} even under the dense CSI dataset in outdoor scenarios. This verifies the effectiveness of the proposed IdasNet for CSI reconstruction not only for sparse channels but also for dense channels.

\begin{figure}[t]
\centering
\includegraphics[scale=0.14]{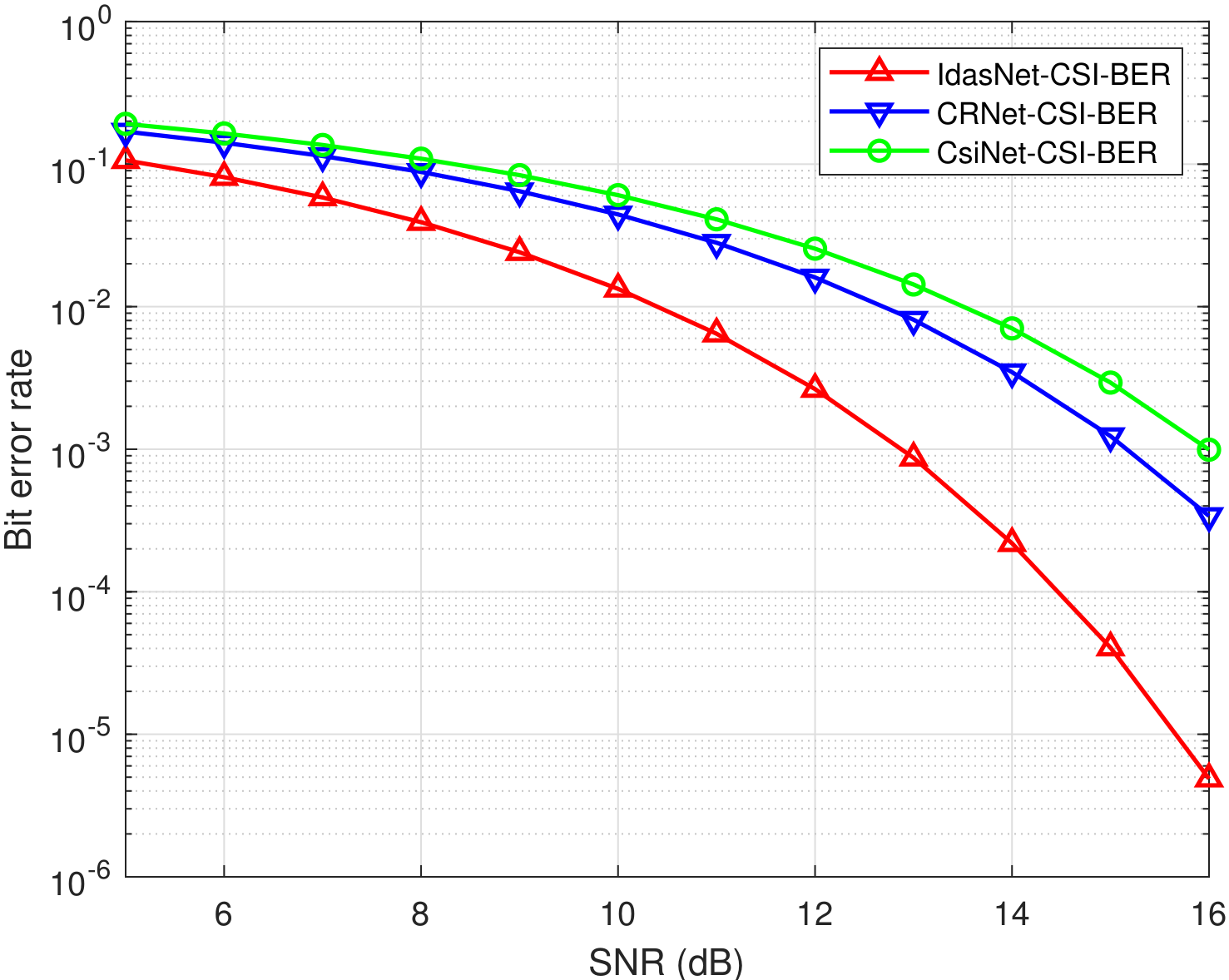}
\caption{BER performance comparison by different reconstructed CSI.}
\label{fig:label11}
\end{figure}

\subsection{Distribution Visualization of Self-information}
% To verify the concept that the codeword values which are acquired by the CSI image without informative redundancy are more conducive to accurate CSI reconstruction at the BS, in Fig. 10 
We visualize the data distribution before compression in IdasNet and other DL-based networks in Fig. 10. We randomly and equally sample various data points from the input of existing encoder and the input of IFC encoder. The data distribution of the existing encoder input in DL networks is shown in Fig. 10(a). The horizontal axis denotes value of elements, and the vertical axis denotes the number of elements. From Fig. 10(a), the data presents a Gaussian-like distribution. The compression encoder using FC layer ignores the structural features of pixels in CSI image, and the essential information and the informative redundancy are compressed into the codeword with an equal probability, which affects the quality of the reconstructed CSI image.

Fig. 10(b) shows the data distribution of input of the proposed IFC encoder. We observe that the data presents a long tail distribution. During the compression, the IFC encoder selects the elements with large self-information value as codeword values, which ensures that the codeword contains more essential information. The codeword with more essential information is conductive to reduce the error of CSI reconstruction, especially under the low compression ratio. Otherwise, long tail distribution has lower entropy than Gaussian-like distribution, which helps the encoder compress the image better. 

% The comparison of data distribution in Fig. 10 illustrates how the proposed IdasNet achieve better performance for CSI feedback than other DL-based methods, especially under the lower compression ratio.

\subsection{Bit Error Rate for Reconstructed CSI}
In this section, we discuss the relation between the NMSE performance of reconstructed CSI and the BER performance. Actually, the CSI with less NMSE implies better BER performance. In general, smaller NMSE represents the lower BER (i.e., better communication performance). This is generally true as accurate channel information is always beneficial to the optimal precoding design at the BS.

Specifically we simulate the BER performance using the reconstructed CSI by different DL-based networks, such as CsiNet \cite{8322184}, CRNet \cite{9149229}, and proposed IdasNet in Fig. 11. The compression ratio of simulation is set to $1/16$ for an instance. The Quadrature Phase Shift Keying (QPSK) modulation is adopted and the precoding vector is a maximum-ratio-transmission (MRT) beamforming designed by using the reconstructed CSI.  From Fig. 11 and Fig. 7(b), it is observed that the proposed IdasNet achieves the smallest NMSE and also significantly outperforms the other methods in terms of BER under different signal-to-noise ratios (SNRs). It validates the above claim that less NMSE indicates a lower BER.

\begin{table}[t]
\centering
\caption{NMSE Comparison with Different Numbers of Neighboring Patches}
\resizebox{100mm}{18mm}{
\begin{tabular}{@{}ccc@{}}
\toprule
Number of neighboring patches               & Compression ratio                          & NMSE (dB) \\ \midrule
\multicolumn{1}{c|}{9 neighboring patches}  & \multicolumn{1}{c|}{\multirow{3}{*}{1/16}} & -13.51    \\
\multicolumn{1}{c|}{27 neighboring patches} & \multicolumn{1}{c|}{}                      & -13.59    \\
\multicolumn{1}{c|}{49 neighboring patches} & \multicolumn{1}{c|}{}                      & -13.68    \\ \midrule
\multicolumn{1}{c|}{9 neighboring patches}  & \multicolumn{1}{c|}{\multirow{3}{*}{1/64}} & -9.34     \\
\multicolumn{1}{c|}{27 neighboring patches} & \multicolumn{1}{c|}{}                      & -9.38     \\
\multicolumn{1}{c|}{49 neighboring patches} & \multicolumn{1}{c|}{}                      & -9.46     \\ \bottomrule
\end{tabular}}
\end{table}

\subsection{Impact of Different Numbers of Neighboring Patches}
For the IDAS module as shown in Fig. 4, we set the Manhattan radius as $R=3$ for the calculation of self-information. Note that when the Manhattan radius $R$ is set to 3, there are 49 neighboring patches for each patch $\textbf{p}_j$. If the IDAS module exploits all the 49 neighboring patches to calculate the self-information value of $\textbf{p}_j$, the calculation workload is prohibitive. To overcome this complexity issue, we adopt a part of the $49$ neighboring patches, i.e., $9$ neighboring patches, which is verified by intensive numerical experiments that they achieve only marginal NMSE loss while reducing the calculation workload greatly. To verify the above statements, we compare the NMSE performance by using different numbers of neighboring patch for $\textbf{p}_j$ under different compression ratios, which is shown in Table \uppercase\expandafter{\romannumeral5}. Note that we choose the number of neighboring patch for $\textbf{p}_j$ as $9$, $27$, and $49$.

It is observed that the NMSEs are quite close between using $9$ neighboring patches and $49$ neighboring patches, but the amount of calculation is reduced by nearly five times when calculating the self-information by using only $9$ neighboring patches for every $\textbf{p}_j$. Thus, in our paper, we select $9$ neighboring patches centered at every $\textbf{p}_j$ instead of all the neighboring patches.

%% file: sections/05_conclusion.tex
\section{Conclusion}
This paper proposed a model-and-data-driven network for CSI compression and feedback. The existing DL methods considered the CSI matrix as an image, but ignored the structural features of the image. Based on this observation, we first proposed a model of self-information to extract the structural features of the CSI image, and introduced the calculation of self-information. In particular, in the proposed network, we designed an informative model-driven module of self-information deletion and selection, referred to as IDAS module. This module pre-compressed, i.e., removing informative redundancy, the CSI image based on self-information. Furthermore, we designed an encoder of informative feature compression, which compressed the CSI image after removing informative redundancy to a codeword according to self-information values. Then, we designed a decoder of informative feature recovery to reconstruct the CSI image at the BS. The experimental results showed that the proposed network outperformed existing DL-based networks for CSI compression and feedback in terms of reconstruction accuracy, especially under the low compression ratio, and required only less network complexity. Additionally, establishing an analytical framework to serve as a baseline for DL-based CSI feedback is an essential task to be addressed in the future work.